\documentclass[conference]{IEEEtran}
\IEEEoverridecommandlockouts
\usepackage{cite}
\usepackage{amsmath,amssymb,amsfonts}
\usepackage{graphicx,color,overpic,psfrag}
\usepackage{xcolor}
\usepackage{algorithmic}
\usepackage{graphicx}
\usepackage{textcomp}
\usepackage{fancyhdr}
\usepackage{xcolor}
\usepackage{graphicx}
\usepackage{bm}
\usepackage{subfigure}
\usepackage{multirow}
\usepackage{booktabs}
\usepackage{caption}
\usepackage{booktabs}
\usepackage{amsmath}
\usepackage[numbers,sort&compress]{natbib}
\usepackage{algorithm}
\usepackage{algorithmic}
\usepackage{amsfonts}
\usepackage{hyperref}
\usepackage{xcolor}
\usepackage{tabularx}
\hypersetup{
  colorlinks=true,
  linkcolor=blue,  
  citecolor=blue,  
  urlcolor=blue,   
  linkbordercolor={0 0 1},
  pdfborderstyle={/S/U/W 1} 
}
\usepackage{booktabs}
\usepackage{makecell}

\DeclareMathOperator*{\argmax}{arg\,max}

\def\BibTeX{{\rm B\kern-.05em{\sc i\kern-.025em b}\kern-.08em
    T\kern-.1667em\lower.7ex\hbox{E}\kern-.125emX}}

\begin{document}

\title{Structure-Aware Multimodal LLM Framework for Trustworthy Near-Field Beam Prediction

\author{Mengyuan Li, \textit{Graduate Student Member, IEEE}, Qianfan Lu, Jiachen Tian, Hongjun Hu, \\Yu Han, \textit{Senior Member, IEEE}, Xiao Li, \textit{Senior Member, IEEE}, \\ Chao-Kai Wen, \textit{Fellow, IEEE}, and Shi Jin, \textit{Fellow, IEEE}\\
}
\thanks{M. Li, Q. Lu, J. Tian, H. Hu, Y. Han, X. Li, and S. Jin are with the School of Information Science and Engineering, Southeast University, Nanjing 210096, China (email: mengyuan$\_$li@seu.edu.cn; qianfan$\_$lu@seu.edu.cn; tianjiachen@seu.edu.cn; huhongjun@seu.edu.cn; hanyu@seu.edu.cn; li$\_$xiao@seu.edu.cn; jinshi@seu.edu.cn). 

C.-K. Wen is with the Institute of Communications Engineering, National Sun Yat-sen University, Kaohsiung 804, Taiwan (e-mail: chaokai.wen@mail.nsysu.edu.tw).
}
}
\maketitle

\begin{abstract}
In near-field extremely large-scale multiple-input multiple-output (XL-MIMO) systems, spherical wavefront propagation expands the traditional beam codebook into the joint angular-distance domain, rendering conventional beam training prohibitively inefficient, especially in complex 3-dimensional (3D) low-altitude environments. Furthermore, since near-field beam variations are deeply coupled not only with user positions but also with the physical surroundings, precise beam alignment demands profound environmental understanding capabilities. To address this, we propose a large language model (LLM)-driven multimodal framework that fuses historical GPS data, RGB image, LiDAR data, and strategically designed task-specific textual prompts. By utilizing the powerful emergent reasoning and generalization capabilities of the LLM, our approach learns complex spatial dynamics to achieve superior environmental comprehension. To mitigate the curse of near-field codebook dimensionality, we design a structure-aware beam prediction head. By decoupling the high-dimensional beam index into independent azimuth, elevation, and distance components, our approach explicitly mirrors the intrinsic 3D geometry of the near-field codebook, enhancing physical interpretability and effectively guiding the learning process. Meanwhile, an auxiliary trajectory prediction head acts as a spatial prior to guide the beam search. Furthermore, to ensure trustworthy prediction against model uncertainties, the framework concurrently outputs confidence scores to trigger an adaptive refinement mechanism, balancing beam alignment accuracy and pilot overhead. Extensive evaluations demonstrate that our framework significantly outperforms state-of-the-art deep learning (DL)-based prediction algorithms and efficient near-field beam training baselines in both line-of-sight and non-line-of-sight scenarios, with rigorous ablation studies confirming the effectiveness of each proposed module in the framework.
\end{abstract}


\section{Introduction} 

\IEEEPARstart{E}{x}tremely large-scale multiple-input multiple-output (XL-MIMO) has emerged as a key technology for sixth-generation (6G) wireless systems~\cite{wang2024xlmimo}. By equipping the base station (BS) with hundreds or thousands of antennas, XL-MIMO significantly enlarges the array aperture, enabling high spatial resolution and substantial array gain, thereby enhancing spectral efficiency and supporting ultra-high data rates and seamless coverage~\cite{han2023toward,liu2025near}. However, the enlarged aperture fundamentally alters the propagation regime. In near-field XL-MIMO systems, spherical wavefront propagation replaces the conventional planar-wave assumption, coupling angular and distance dimensions and producing volumetric beam patterns~\cite{cui2022channel}. As a result, near-field beams become extremely narrow and highly position-sensitive. Although high directivity improves energy focusing, it also increases sensitivity to misalignment, leading to severe degradation of achievable rates~\cite{luo2024efficient}. Consequently, efficient beam alignment becomes a structural requirement in near-field XL-MIMO systems.

\subsection{Prior Works}

Beam management has been extensively investigated to reduce training overhead while maintaining reliable alignment. Existing works can be broadly classified into two primary paradigms: beam training-based methods and beam prediction-based methods. The former improves search efficiency through structured pilot sweeping, whereas the latter aims to directly infer future optimal beams from historical observations. More recently, beam prediction has evolved toward multimodal environment-aware learning, incorporating heterogeneous sensing information to improve semantic awareness and robustness.

\subsubsection{Beam Training-Based Methods}
Conventional beam training determines the optimal beam via pilot sweeping. Although reliable, exhaustive search incurs substantial overhead and suffers from information aging in dynamic scenarios~\cite{xu2025near}. Numerous studies have therefore proposed structured search strategies to reduce pilot consumption.
In far-field systems, various hierarchical and adaptive schemes are designed to effectively balance beam training overhead and accuracy~\cite{shokri2015beam,yaman2016reducing,shen2018mobility,qi2020hierarchical}. However, in near-field XL-MIMO systems, spherical wavefront propagation introduces an additional distance dimension, causing exponential growth of the angular-distance codebook~\cite{cui2023near,li2025keypoint}. The resulting high-dimensional volumetric search space significantly increases computational and signaling complexity.
To mitigate this issue, hierarchical and multi-stage strategies progressively refine angular and distance estimates~\cite{lu2024hierarchical,wu2024two}. Other approaches exploit structural properties of beam patterns to reuse far-field discrete Fourier transform codebooks for near-field estimation~\cite{wu2024near}. From a probabilistic viewpoint, Bayesian regression frameworks model codeword correlations to infer optimal beams with limited measurements~\cite{xu2025near}.  
Despite these advances, beam training remains fundamentally a measurement-driven search procedure. As codebook dimensionality and mobility increase, pilot overhead and latency become increasingly prohibitive in practical near-field deployments.

\subsubsection{Wireless-Only Beam Prediction}
To further reduce online search complexity, beam prediction methods aim to forecast optimal beams directly from historical wireless observations~\cite{chen2023mmwave,jayaprakasam2017robust,khunteta2021recurrent,shah2022multi,lin2025bridge,xue2024ai}. Early approaches rely on kinematic models and sequential estimation techniques such as the extended Kalman filter and particle filter~\cite{chen2023mmwave,jayaprakasam2017robust}. However, their reliance on simplified motion assumptions limits robustness under nonlinear mobility and rapid channel variation.
With the advancement of artificial intelligence, data-driven models have emerged as a dominant paradigm ~\cite{lin2025bridge,xue2024ai}. Recurrent neural networks (RNNs) and long short-term memory (LSTM) networks have been widely adopted to learn temporal correlations between historical beams or pilot signals and future optimal beams~\cite{khunteta2021recurrent,shah2022multi}. By formulating beam prediction as a sequence learning problem, these methods effectively capture nonlinear temporal dynamics and significantly reduce search overhead.

Nevertheless, most existing prediction approaches rely solely on wireless measurements. In near-field XL-MIMO systems, optimal beam selection is intrinsically coupled with user position and surrounding geometry due to volumetric beam characteristics. The exponential growth of angular-distance codebooks expands the prediction output space beyond conventional classification scalability. Without explicit geometric and environmental semantics, wireless-only models face fundamental generalization limitations.

\subsubsection{Multimodal Environment-Aware Beam Prediction}
To address the lack of environmental awareness, recent studies incorporate heterogeneous sensing modalities such as RGB images, LiDAR point clouds, and GPS data into beam prediction frameworks~\cite{charan2022vision,charan2024camera,jiang2023lidar,zhao2025multi,zheng2025m2beamllm,sheng2025beam,liu2025large}. By explicitly modeling scatterer distributions and blockage conditions, multimodal approaches enhance robustness in complex propagation environments.
Motivated by the success of generative AI, recent works have increasingly explored large language models (LLMs) to process multimodal information for beam prediction~\cite{liu2025large,zhao2025multi,zheng2025m2beamllm,sheng2025beam}. This paradigm shift is driven by three intrinsic advantages: (i) massive pre-training provides superior generalization across diverse communication scenarios; (ii) task-specific prompts significantly enhance the model's comprehension of customized prediction objectives; and (iii) LLMs exhibit exceptional capacity in fitting high-dimensional heterogeneous data.
Specifically, MLM-BP~\cite{zhao2025multi} adopts a DeepSeek-based multimodal model~\cite{chen2025janus} to fit scatterer distributions tokenized by a LoRA-tuned image encoder~\cite{hu2022lora}. To capitalize on task-specific prompting, \cite{sheng2025beam} employs a prompt-as-prefix strategy to encode historical beams and environmental states, effectively reformulating beam prediction as a language reasoning task to guide the LLM's understanding. Building upon these foundational strengths, M2BeamLLM~\cite{zheng2025m2beamllm} performs rigorous multimodal feature alignment before LLM inference, fully unlocking the pre-trained model's capacity to process and fit different modality features within a unified semantic space.

However, multimodal near-field beam prediction still faces several critical challenges. First, most existing studies still focus on far-field propagation, which is largely restricted by the attributes of prevalent benchmarks, such as DeepSense 6G~\cite{alkhateeb2023deepsense} and Multimodal-Wireless~\cite{mao2025multimodal} that are tailored for far-field settings, failing to capture the spherical wavefronts and the unique spatial-selective characteristics in near-field regions. Second, the exponential expansion of joint angular-distance codebooks renders direct codeword-level classification inefficient and poorly scalable in 3-dimensional (3D) low-altitude environments, necessitating the structural exploitation of geometric constraints. Third, current multimodal frameworks prioritize accuracy while overlooking reliability. The absence of confidence assessment and adaptive fallback mechanisms inevitably leads to unstable system performance in high-mobility scenarios.

\subsection{Main Contributions}
To address the aforementioned challenges, we propose a structure-aware multimodal LLM framework for trustworthy beam prediction in near-field XL-MIMO systems. Specifically, our main contributions are summarized as follows:

\begin{itemize}

\item \textbf{Multimodal Inputs and LLM Reasoning:} 
We design tailored encoders to effectively extract rich environmental semantics from diverse multimodal inputs. By fusing these representations with proposed task- and trajectory-related textual prompts, it provides contextual guidance that empowers the LLM backbone. This fully unleashes the emergent reasoning capabilities of LLM to achieve a profound understanding of the environment. Since the optimal near-field beam index is highly coupled with physical spatial geometries, this deep environmental awareness significantly boosts the accuracy.

\item \textbf{Structure-Aware Beam Prediction with Auxiliary Trajectory Guidance:}
To mitigate the curse of dimensionality, we utilize a structure-aware, decoupled prediction strategy. By independently predicting the azimuth, elevation, and distance indices, our approach explicitly mirrors the near-field codebook's intrinsic 3D geometry, which effectively guides the learning process and significantly boosts prediction accuracy. Furthermore, we introduce an auxiliary trajectory prediction head to capture the UAV's future motion dynamics, which acts as a spatial prior to guide the beam search process and to further improve the accuracy.

\item \textbf{Confidence-Aware Adaptive Refinement:}
To combat model uncertainties, we propose a confidence-driven strategy that dynamically triggers small-scale beam scanning within the predicted candidate pool only when the confidence score is low. This mechanism optimally balances pilot overhead and highly accurate beam alignment.

\item \textbf{Comprehensive Validation and Ablation Studies:}
Extensive experiments under both LoS and NLoS conditions demonstrate consistent performance gains over state-of-the-art (SOTA) sequence prediction models and efficient near-field beam training baselines. Furthermore, rigorous ablation studies are conducted to validate the necessity and effectiveness of each core component within the proposed framework.

\end{itemize}

{\bf Notations.} Bold uppercase and lowercase letters denote matrices and vectors, respectively. $(\cdot)^{\top}$ and $(\cdot)^\mathsf{H}$ denote transpose and conjugate transpose. $\odot$, $\mathbb{E}\{\cdot\}$, $|\cdot|$, and $\|\cdot\|$ represent the Hadamard product, expectation, absolute value, and Euclidean norm, respectively. $\mathbf{I}$ denotes the identity matrix. $\mathcal{CN}$ represents the complex Gaussian distribution. $\mathbb{I}(\cdot)$ denotes the indicator function, which equals 1 if the condition holds and 0 otherwise. Finally, $\mathrm{clamp}(x, a, b) = \max(a, \min(x, b))$ restricts the value of $x$ to the interval $[a, b]$.

\section{System Model}
\label{sec:System Model}

In this section, we first establish the near-field channel and XL-MIMO system models. Building upon these foundations, the beam prediction task is formulated as a sequential multimodal prediction problem.

\subsection{Channel Model}

\begin{figure}[t]
\centering
\includegraphics[width=0.98\linewidth]{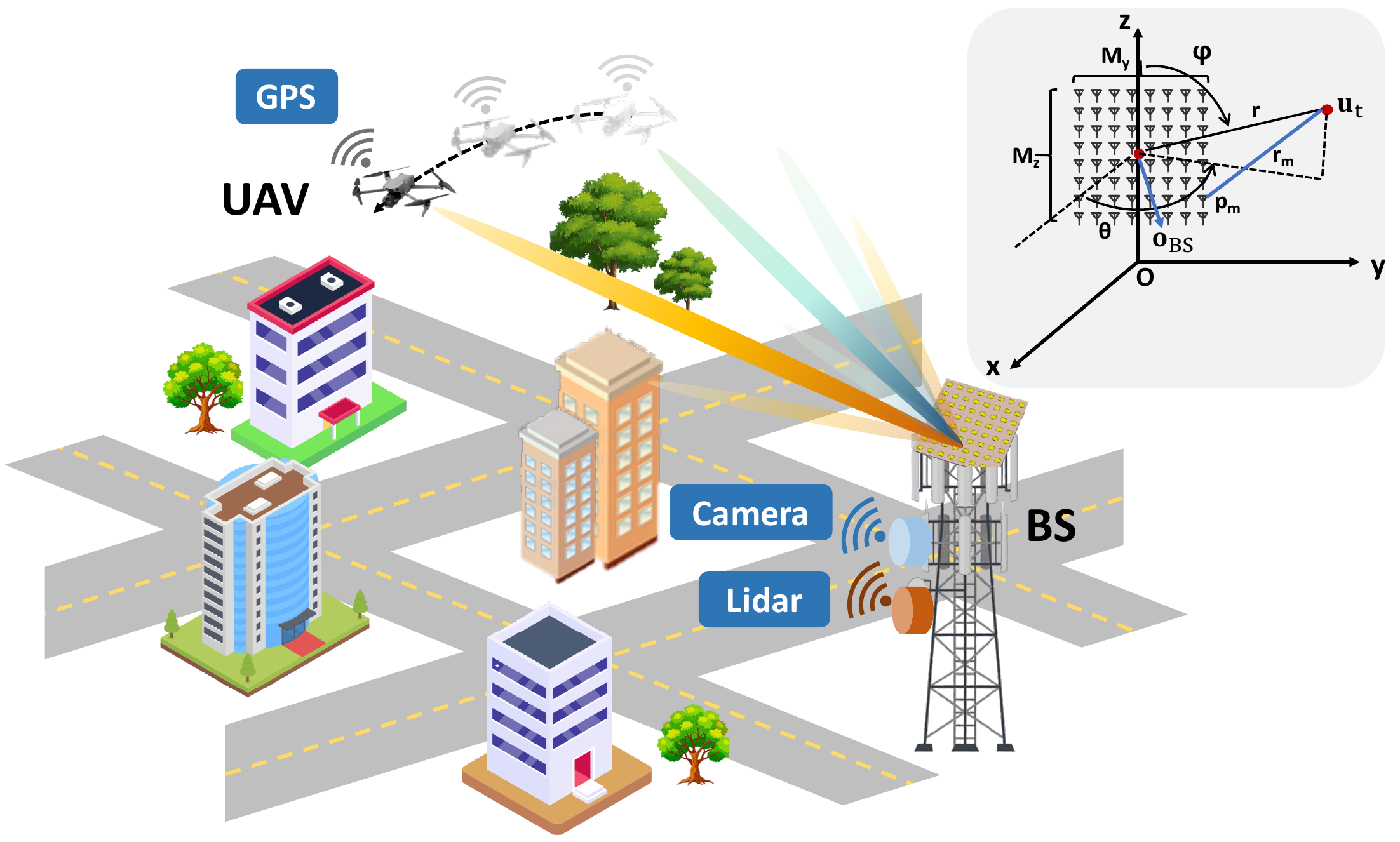}
\caption{Illustration of the XL-MIMO system model in LAE scenarios: The BS is equipped with a UPA, an RGB camera, and a LiDAR, while the UAV is equipped with a GPS which feeds back locations to the BS.}

\label{fig:LAE system}
\end{figure}

We consider a single-cell XL-MIMO system operating in the urban low-altitude environment, as illustrated in Fig.~\ref{fig:LAE system}. The system consists of a BS and a mobile UAV as the user equipment (UE). To facilitate multi-modal environment perception, the BS is equipped with an RGB camera and a LiDAR sensor alongside a uniform planar array (UPA) with $M = M_y \times M_z$ antennas, where $M_y$ and $M_z$ denote the number of antenna elements along the horizontal and vertical axes, respectively. The antenna element spacing is set as $d_y = d_z = 0.5\lambda$, where $\lambda = c/f_c$ is the wavelength at carrier frequency $f_c$, and $c$ is the speed of light. The position of the $m$-th antenna element, indexed by $(m_y, m_z)$, is given by
\begin{equation}
    \mathbf{p}_{m} 
    = \mathbf{o}_\text{BS} + 
    \Big[ 0, \; \big(m_y - \tfrac{M_y-1}{2}\big)d_y, \; \big(\tfrac{M_z-1}{2} - m_z\big)d_z \Big]^{\mathsf{T}},
\end{equation}
where $\mathbf{o}_\text{BS}$ denotes the center of the UPA, $m_y \in \{0,\ldots,M_y-1\}$ and $m_z \in \{0,\ldots,M_z-1\}$.
The UAV is assumed to operate within the near-field region. Meanwhile, the UAV is equipped with an onboard GPS receiver. It is equipped with a single omnidirectional antenna and follows a time-varying 3D trajectory, with its instantaneous location at time $t$ denoted by $\mathbf{u}_t \in \mathbb{R}^3$.

We employ the Sionna ray tracing (RT)~\cite{sionnaRT} for high-fidelity near-field channel generation. It computes the channel impulse responses by combining shooting-and-bouncing rays (SBR) with the image method, simulating the physical interaction of wavefronts with environmental scatterers.
The time-varying near-field uplink channel vector $\mathbf{h}(t) \in \mathbb{C}^{M \times 1}$ is composed of the channel response $h_m(t)$ for each receive antenna $m$. Specifically, $h_m(t)$ is modeled as:
\begin{equation}
    \label{eq:channel_model}
    h_{m}(t) 
    = \sum_{l=1}^{L(t)} g_{l,m}(t) \, e^{-j \frac{2\pi}{\lambda} d_{l,m}(t)},
\end{equation}
where $L(t)$ is the number of propagation paths, $g_{l,m}(t)$ and $d_{l,m}(t)$ represent the complex path gain and the propagation path length of the $l$-th path arriving at the $m$-th antenna, respectively.
Unlike the far-field plane wave assumption, the path length $d_{l,m}(t)$ is calculated based on the specific propagation topology and the exact Euclidean distance to each antenna element.
For the line-of-sight (LoS) path, the distance can be expressed as
\begin{equation}
d_{l,m}(t) = \|\mathbf{u}_t - \mathbf{p}_{m}\|.
\end{equation}
For non-LoS (NLoS) paths, $d_{l,m}(t)$ is geometrically calculated as the sum of physical distances between consecutive interaction points (e.g., reflections, diffractions, or scattering). 
Furthermore, the complex gain $g_{l,m}(t)$ explicitly captures the path loss, antenna polarization matching, and electromagnetic (EM) material properties. Instead of relying on simplified statistical formulations, $g_{l,m}(t)$ is deterministically computed using the Sionna ray tracer (RT)~\cite{sionnaRT}, which accurately evaluates the EM transfer matrices and spatial field patterns along the precise trajectory of each ray.

Assuming the UAV transmits pilot symbols with power $P_r$, the received signal vector $\mathbf{y}(t) \in \mathbb{C}^{M\times1}$ at the BS can be expressed by
\begin{equation}
    \mathbf{y}(t) = \sqrt{P_r}\,\mathbf{w}^{\mathsf{H}}\mathbf{h}(t) + \mathbf{w}^{\mathsf{H}}\mathbf{n}(t),
\end{equation}
where $\mathbf{w} \in \mathbb{C}^{M\times1}$ is the beamforming vector, $\mathbf{n}(t) \sim \mathcal{CN}(\mathbf{0}, \sigma^2\mathbf{I})$ is the additive white Gaussian noise.

\subsection{Problem Formulation}

We first construct a polar-domain codebook $\mathcal{W}$ by jointly sampling the angular and distance domains:
\begin{equation}
    \mathcal{W} = \big\{ \mathbf{w}(\theta_i,\varphi_j,r_q) \mid 1 \le i \le N_{\theta}, \, 1 \le j \le N_{\varphi}, \, 1 \le q \le N_{r} \big\},
\end{equation}
where $N_{\theta}, N_{\varphi},$ and $N_{r}$ denote the number of sampled codewords for the azimuth angle, elevation angle, and distance, respectively. The near-field codeword corresponding to the tuple $(\theta_i, \varphi_j, r_q)$ is defined as
\begin{equation}
    \mathbf{w}(\theta_i,\varphi_j,r_q) = \frac{1}{\sqrt{M}} \big[ e^{-j \frac{2\pi}{\lambda} (\|\mathbf{p}_{\text{cw}} - \mathbf{p}_{1}\|)}, \ldots, e^{-j \frac{2\pi}{\lambda} (\|\mathbf{p}_{\text{cw}} - \mathbf{p}_{M}\|)} \big]^{\mathsf{T}},
\end{equation} 
where $\mathbf{p}_{\text{cw}}$ is the Cartesian coordinate of the sampled point.  

For a selected beam codeword $\mathbf{w} \in \mathcal{W}$ at time slot $t$, the achievable rate is defined as
\begin{equation} 
    R(\mathbf{w}, t) = \log_2\!\left( 1 + \frac{P_r \left|\mathbf{w}^{\mathsf{H}}\mathbf{h}(t)\right|^2}{\sigma^2}\right),
\label{eq:achievable_rate}
\end{equation}
where $P_r$ is the transmit power and $\sigma^2$ is the noise variance. The objective of beam management is to select the optimal codeword $\mathbf{w}^\star$ that maximizes~\eqref{eq:achievable_rate}, which is equivalent to maximizing the received beamforming gain. Specifically, we first define the beamforming gain for a given codeword $\mathbf{w}$ at time slot $t$ as:
\begin{equation}
    G(\mathbf{w}, t) = \left|\mathbf{w}^\mathsf{H}\mathbf{h}(t)\right|^2.
\label{eq:beam_gain}
\end{equation}
The optimal codeword is then obtained by finding the maximum gain across the codebook:
\begin{equation}
    \mathbf{w}^\star = \argmax_{\mathbf{w}\in\mathcal{W}} G(\mathbf{w}, t). 
\label{eq:optimal_beam_gain} 
\end{equation}
While exhaustive beam search guarantees optimal selection, evaluating the massive near-field codebook of size $N_{\theta} N_{\varphi} N_{r}$ incurs prohibitively high training overhead and latency. 

To bypass the exhaustive search overhead, we formulate beam training as a sequential prediction task. Instead of relying solely on traditional pilot signals, we design multimodal encoders to extract a unified representation $\mathcal{E}_t$ that encapsulates the environmental context from sensing data alongside the UAV's kinematic information over a historical window $[t-L_h, t]$. 
Furthermore, predicting a single global beam index over the massive near-field codebook creates an unwieldy action space that severely hinders network convergence. To accelerate the learning process and reduce complexity, we decouple the prediction across the three spatial dimensions. Our objective is to learn a mapping function $\mathcal{F}_{\Theta}(\cdot)$ that directly predicts the optimal decoupled index triplets for a subsequent horizon of length $L_{p}$:
\begin{equation}
    \big[ (\hat{i}, \hat{j}, \hat{q})_{t+1}, \dots, (\hat{i}, \hat{j}, \hat{q})_{t+L_p} \big] = \mathcal{F}_{\Theta} \big( \mathcal{E}_t \big),
\end{equation}
where $(\hat{i}, \hat{j}, \hat{q})_{\tau}$ denotes the predicted sub-indices for azimuth, elevation, and distance, respectively, at future time slot $\tau \in \{t+1, \dots, t+L_p\}$.

To establish the ground truth for our predictive model, we define the optimal sub-indices for azimuth $i^\star$, elevation $j^\star$, and distance $q^\star$ at time slot $t$ by jointly maximizing the beamforming gain:
\begin{equation}
    (i^\star, j^\star, q^\star) = \argmax_{\substack{1 \le i \le N_{\theta} \\ 1 \le j \le N_{\varphi} \\ 1 \le q \le N_{r}}} \left|\mathbf{w}(\theta_i,\varphi_j,r_q)^\mathsf{H}\mathbf{h}(t)\right|^2.
\label{eq:optimal_sub_indices}
\end{equation}
The overall optimal beam index $k^\star \in \{1, \dots, N_{\theta} N_{\varphi} N_{r}\}$ can be uniquely mapped from this triplet via
\begin{equation}
    k^\star = (i^\star - 1) N_{\varphi} N_{r} + (j^\star - 1) N_{r} + q^\star.
\label{eq:optimal_global_index}
\end{equation}

\section{Structure-Aware Multimodal LLM Framework} \label{sec:LLM-Driven Multimodal-Aware Beam Prediction Framework}
In this section, we elaborate on the proposed structure-aware LLM-driven multimodal beam prediction framework. We first present an overview of the proposed framework. Then, we detail the core components, including multimodal encoders and the feature fusion module, the structure-aware beam prediction head, and the adaptive refinement mechanism. Finally, we describe the training scheme and loss function design.

\subsection{Overall Workflow}

The overall workflow of the proposed framework is shown in Fig.~\ref{fig:workflow}. It adopts a \textit{``representation-perception-fusion-reasoning-refinement''} paradigm through the following five modules:

\begin{figure*}[t]
\centering
\includegraphics[width=0.98\linewidth]{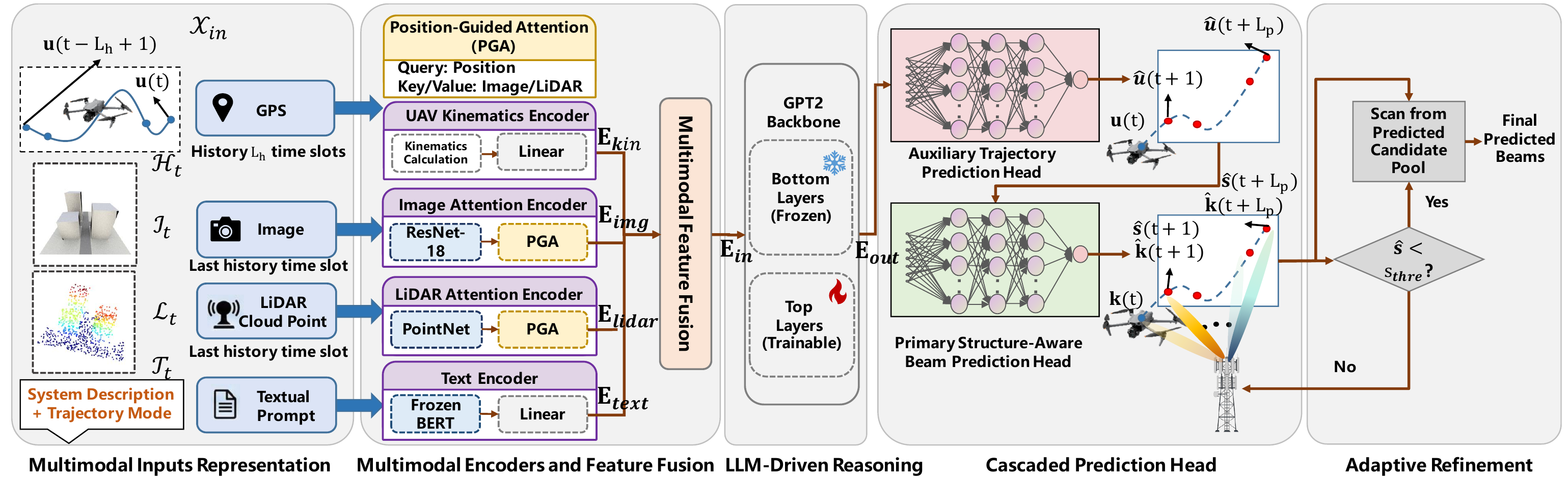}
\caption{Overall workflow of the proposed structure-aware LLM-driven multimodal beam prediction framework.}
\label{fig:workflow}
\end{figure*}

\subsubsection{Multimodal Input Representation}
To accurately predict the optimal beam index by capturing the complex interplay between the UAV's kinematic state and the wireless propagation environment, we first formulate the multimodal input set $\mathcal{X}_\text{in} = \{ \mathcal{H}_t, \mathcal{I}_t, \mathcal{L}_t, \mathcal{T}_t \}$, which integrates complementary information across distinct modalities, including:

\begin{itemize}
    \item \textbf{Historical Kinematics ($\mathcal{H}_t$):} To capture the UAV's temporal motion trajectory, we construct a sequence of historical positions $\mathcal{H}_t = \{\mathbf{u}(\tau)\}_{\tau=t-L_h+1}^{t}$, where $\mathbf{u}(\tau) \in \mathbb{R}^{N\times 1\times 3}$ denotes the 3D coordinate at time slot $\tau$, acquired via an onboard GPS receiver and subsequently fed back to the BS. $N$ is the batch size. Specifically, we model the measurement error of GPS by $\mathbf{u}(\tau) = \tilde{\mathbf{u}}(\tau) + \mathbf{n}(\tau)$, where $\tilde{\mathbf{u}}(\tau)$ is the true UAV position and $\mathbf{n}(\tau) \sim \mathcal{N}(\mathbf{0}, \sigma_{\text{GPS}}^2 \mathbf{I})$ represents the additive Gaussian noise with standard deviation $\sigma_{\text{GPS}}$.

    \item \textbf{Visual and Depth Data ($\mathcal{I}_t, \mathcal{L}_t$):} To comprehensively perceive the environment, both an RGB camera and a LiDAR are deployed at the BS. The camera provides RGB images $\mathcal{I}_t$ containing texture and blockage information, whereas the LiDAR generates point clouds $\mathcal{L}_t$ detailing the precise depth and geometric structure of the scattering environment. To avoid the memory overhead and processing latency associated with sequence modeling, the proposed scheme relies solely on the instantaneous sensory observations $\mathcal{I}_t$ and $\mathcal{L}_t$ at current time slot $t$.
    
    \item \textbf{Textual Prompts ($\mathcal{T}_t$):} To inject domain knowledge, we construct textual prompts $\mathcal{T}_t$ that encompass static system descriptions (e.g., operating frequency, antenna array size) and dynamic descriptions of the UAV's flight mode (e.g., ``Zigzag'', ``Street Patrol'').

\end{itemize}
These inputs are then fed into their respective encoders to be projected into a unified high-dimensional latent space.

\subsubsection{Multimodal Encoders and Feature Fusion}
The framework first explicitly models the UAV's temporal motion trends by calculating and encoding historical kinematic states from $\mathcal{H}_t$. Simultaneously, to effectively couple the physical environment with the UAV's location, we introduce a position-guided attention (PGA) mechanism that extracts position-related features from RGB images and LiDAR point clouds. Furthermore, semantic guidance is incorporated via a textual prompt encoder that processes system and trajectory descriptions. These multimodal feature streams are ultimately synchronized and concatenated within the following fusion module to form a unified input for the subsequent LLM-driven reasoning backbone.

\subsubsection{LLM-Driven Reasoning}
The fused multimodal features are subsequently passed to a pre-trained GPT-2 model~\cite{radford2019language}\footnote{The pre-trained weights of the GPT-2 model can be found at \url{https://huggingface.co/gpt2}.} for fine-tuning and sequential reasoning. 
Unlike conventional methods that formulate beam prediction as a static classification task, the GPT-2 backbone functions as a context-aware reasoning engine. 
It effectively captures the complex dynamic interactions among the UAV's flight trajectory, the surrounding environmental geometry, and the corresponding optimal beam sequences. 
In this way, the network can deduce how the relative motion between the UAV and physical scatterers (e.g., blockages or reflectors) influences the beam transitions. 
Ultimately, the model establishes a robust spatiotemporal mapping within the latent space, translating historical observations into highly predictive latent representations of future states.

\subsubsection{Cascaded Prediction Heads}
To effectively map the GPT-2 output latent representations to the corresponding wireless channel characteristics, we employ a cascaded dual-head architecture, including:

\begin{itemize}
    \item \textbf{An Auxiliary Trajectory Prediction Head:} The latent representations from the LLM are first processed by an auxiliary network to predict the UAV's future 3D coordinates $\{\widehat{\mathbf{u}}(\tau)\}_{\tau=t+1}^{L_p}$. This trajectory prediction serves as an auxiliary geometric prior rather than the ultimate objective. It forces the latent features to encode the kinematic and surrounding environment evolution, acting as a physical anchor to ground the subsequent beam prediction task.
    
    \item \textbf{A Primary Structure-Aware Beam Prediction Head:} The predicted trajectory is then injected into the primary beam prediction head. By conditioning on the predicted 3D position, the network effectively narrows down the candidate pool, allowing it to ignore geometrically impossible beams and focus solely on environmental features consistent with the UAV's future location. Furthermore, to mitigate the curse of dimensionality associated with the enormous near-field codebook, the prediction head avoids directly estimating a global beam index $\hat{k}$. Instead, it outputs decoupled sub-indices $(\hat{i},\hat{j},\hat{q})$ that independently specify the azimuth, elevation, and distance components of the 3D near-field beam. By decomposing the spatial prediction task, this decoupled scheme acts as a \textbf{structure-aware} predictor that respects the inherent 3D geometry of the near-field codebook. This \textbf{structure-aware} design endows the beam prediction with explicit physical interpretability. By inherently linking the variations in the decoupled sub-indices to the target's actual 3D coordinates in the angular and distance domains, the network avoids the opaque nature of a structureless overall index. This explicit physical grounding effectively guides the learning process, thereby significantly enhancing the prediction accuracy.
\end{itemize}

\subsubsection{Adaptive Refinement Mechanism}
Despite the effectiveness of the proposed network, data-driven predictions inherently exhibit a certain degree of uncertainty. To improve the reliability of beam prediction and guarantee system communication quality with low pilot overhead, we design an adaptive refinement mechanism. Upon generating the beam candidates, the mechanism evaluates the maximum confidence score $\hat{s}$. High-confidence predictions (${\hat{s} > s_{\text{thre}}}$) are accepted immediately for rapid beamforming. In contrast, low-confidence cases (${\hat{s} \leq s_{\text{thre}}}$) activate a targeted refinement process, executing a small-scale beam sweep exclusively among a small-scale beam candidate pool. This selective execution effectively mitigates the impact of model uncertainty, ensuring high-precision tracking while maintaining a significantly lower overhead compared to exhaustive sweeping.

\subsection{Multimodal Encoders and Feature Fusion}

\subsubsection{UAV Kinematics Calculation and Encoding}
To help capture the temporal motion dynamics, we process the UAV's kinematic states over a historical observation window of length $L_h$. The sequence of historical positions is denoted as $\{\mathbf{u}(\tau)\}_{\tau=t-L_h+1}^{t}$. The velocity $\mathbf{v}(\tau)\in \mathbb{R}^{N\times 1 \times 3}$ and acceleration $\mathbf{a}(\tau)\in \mathbb{R}^{N\times 1\times 3}$ are calculated by:
\begin{equation}
    \mathbf{v}(\tau) = \frac{\mathbf{u}(\tau) - \mathbf{u}(\tau-1)}{\Delta t}, \quad \mathbf{a}(\tau) = \frac{\mathbf{v}(\tau) - \mathbf{v}(\tau-1)}{\Delta t},
\end{equation}
where $\Delta t$ is the sampling interval. 
By concatenating these derived states, we construct the historical kinematics sequence $\mathcal{H}_t = \{[\mathbf{u}(\tau), \mathbf{v}(\tau), \mathbf{a}(\tau)]\}_{\tau=t-L_h+1}^{t}$.
This sequence is then projected into the latent space via a learnable linear layer to form the kinematic embedding sequence $\mathbf{E}_\text{kin}\in \mathbb{R}^{N \times L_p \times d_\text{model}}$, which captures the trajectory evolution and serves as the motion context for the beam predictor. $d_\text{model}$ is the unified latent dimension.

\begin{figure}[t]
\centering
\includegraphics[width=0.98\linewidth]{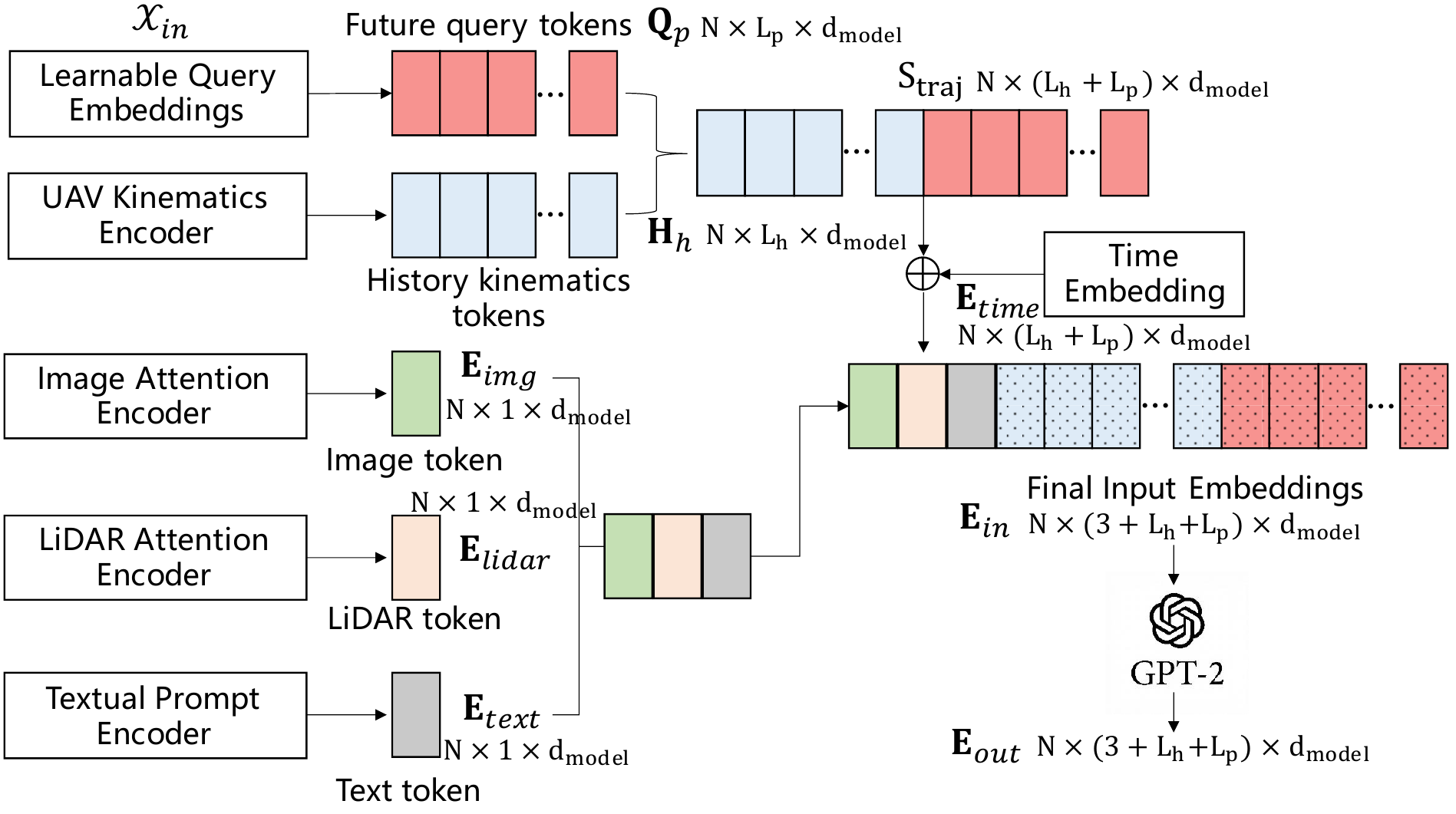}
\caption{Architecture of the designed multimodal feature fusion module.}
\label{fig:feature_fusion}
\end{figure}

\subsubsection{Position-Guided Image and LiDAR Encoders}
To align the multimodal data with the UAV's real-time locations, we introduce a PGA mechanism. As illustrated in Fig.~\ref{fig:workflow} and Fig.~\ref{fig:feature_fusion}, this module serves as a bridge, utilizing the UAV's position $\mathbf{u}(t)$ as a spatial query to actively aggregate high-dimensional sensory features. The detailed mathematical formulation of the PGA cross-attention mechanism is provided in Appendix \ref{appendix:pga}. By explicitly incorporating geometry constraints, the PGA transforms raw inputs into compact, spatially-aware context tokens $\mathbf{E}_\text{img}$ and $\mathbf{E}_\text{lidar}$.

\paragraph{Image Encoder}
We employ a pre-trained ResNet-18~\cite{he2016deep} to extract environmental features from RGB images, such as building footprints and road topologies. The output is flattened to generate the visual feature map $\mathbf{F}_\text{img} \in \mathbb{R}^{N \times 49 \times d_{\text{in}}}$, where $d_{\text{in}}$ denotes the input feature dimension of the PGA module. 
To construct the visual spatial bias $\mathbf{M}_\text{img} \in \mathbb{R}^{N \times 1 \times 49}$, we first project the UAV's 3D coordinate $\mathbf{u}(t)$ onto the 2D image plane using the camera intrinsic parameters. 
Subsequently, we compute $\mathbf{M}_\text{img}$ based on the Gaussian distance between the projected point and the receptive field center of each of the 49 feature tokens. 
This creates a ``soft attention spotlight,'' ensuring that the model inherently prioritizes visual features physically closer to the UAV. 
By feeding $\mathbf{u}(t)$, $\mathbf{F}_\text{img}$, and $\mathbf{M}_\text{img}$ into the PGA module, we obtain the final visual context token $\mathbf{E}_\text{img} \in \mathbb{R}^{N \times 1 \times d_\text{model}}$.
    
\paragraph{LiDAR Encoder} 
Similarly, a PointNet~\cite{qi2017pointnet} backbone processes the point cloud to extract global geometric features. We sample $L_\text{lidar}$ key feature points to obtain the geometric feature map $\mathbf{F}_\text{lidar} \in \mathbb{R}^{N \times 1024\times d_{\text{in}}}$. 
The geometric spatial bias $\mathbf{M}_\text{lidar} \in \mathbb{R}^{N \times 1 \times 1024}$ is directly derived from the 3D Euclidean distance between the UAV coordinate $\mathbf{u}(t)$ and the spatial coordinates of the sampled LiDAR keypoints. 
Guided by this explicit distance bias, the PGA module aggregates the raw point features $\mathbf{F}_\text{lidar}$ into the geometric context token $\mathbf{E}_\text{lidar} \in \mathbb{R}^{N \times 1 \times d_\text{model}}$, thereby heavily weighting the immediate structural constraints surrounding the UAV.

\begin{figure}[t]
\centering
\includegraphics[width=0.98\linewidth]{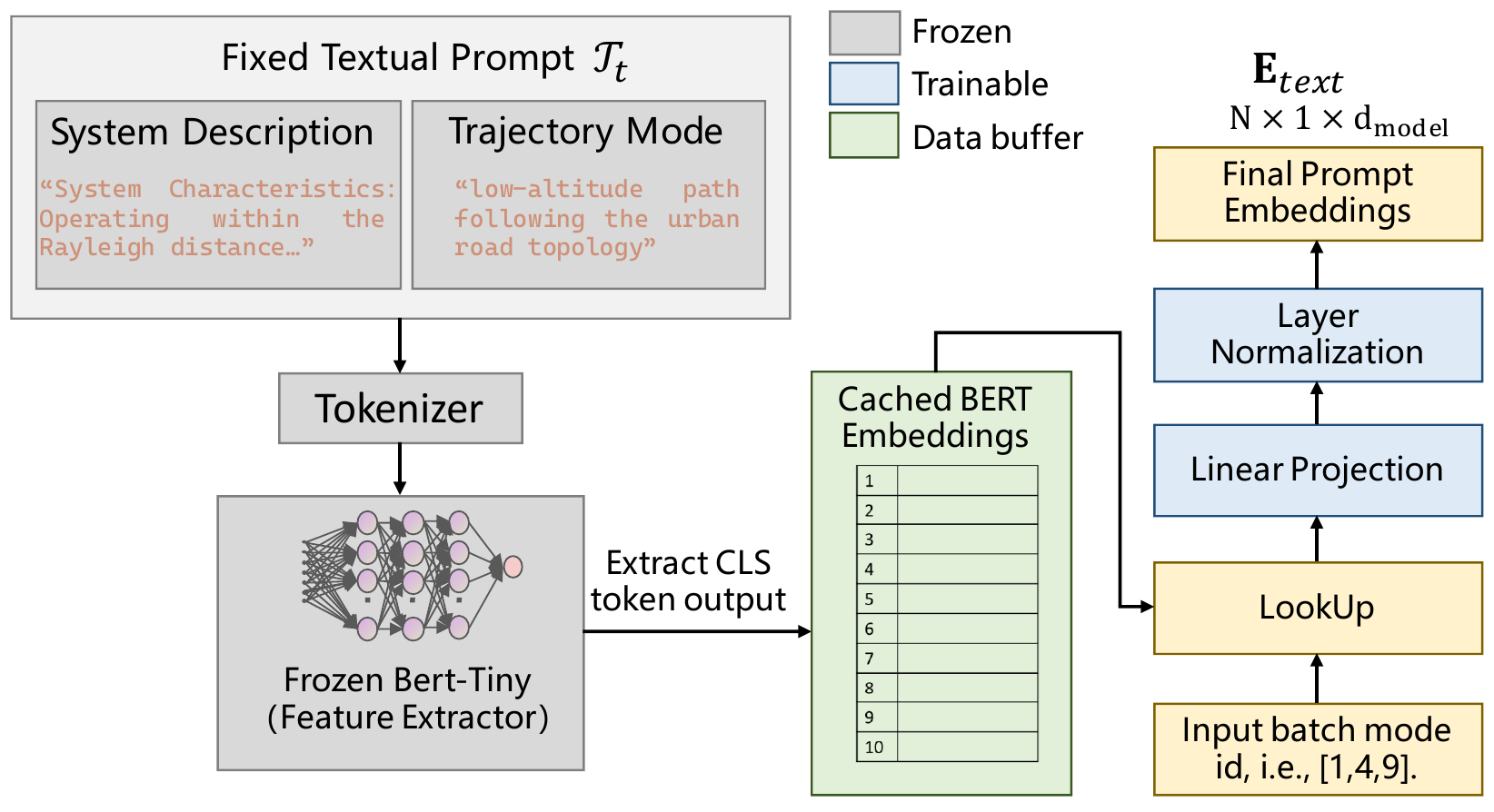}
\caption{Architecture of the designed textual prompt encoder and examples of designed textual prompts.}
\label{fig:text_encoder}
\end{figure}

\subsubsection{Textual Prompt Encoder}
As illustrated in Fig.~\ref{fig:text_encoder}, to efficiently inject high-level semantic guidance, we design a rapid-inference textual encoder that operates on a pre-agreed set of flight modes shared between the BS and the UAV.

\paragraph{Textual Prompt Construction}
To guide the beam prediction, we construct a structured textual prompt $\mathcal{T}_t$ by concatenating two parts: (i) a static \textit{System Description} defining the communication task and the environment; (ii) a dynamic \textit{Trajectory Mode} specifying the current trajectory characteristics (e.g., straight flight or turns).
This textual context $\mathcal{T}_t$ helps the generative model understand the physical intent behind the numerical trajectory data. The detailed prompts are exemplified in Fig.~\ref{fig:text_encoder}.

\paragraph{Offline Caching and Online Lookup}
To reduce real-time latency, we decouple textual prompt encoding from the online inference loop. The prompt consists of a static \textit{System Description} and a dynamic \textit{Trajectory Mode}. Since the trajectory modes fall into predefined categories, we pre-construct all possible prompt combinations offline. A frozen BERT-Tiny~\cite{devlin2019bert} backbone is then utilized to pre-compute their embeddings via the \texttt{[CLS]} token, which are stored in a lightweight look-up table. 
During online inference, the system bypasses expensive text tokenization and encoding. It directly retrieves the pre-computed embedding using the current trajectory mode ID, and projects it via a learnable linear layer to form the global context token $\mathbf{E}_\text{text} \in \mathbb{R}^{N \times 1 \times d_\text{model}}$. This strategy can significantly reduce computational overhead.

\subsubsection{Multimodal Feature Fusion Module}
As illustrated in Fig.~\ref{fig:feature_fusion}, we construct the input sequence by aligning all modalities into a shared latent space. 
First, the historical kinematics sequence $\mathcal{H}_t$ is projected via an MLP to obtain the history token sequence $\mathbf{H}_{h} \in \mathbb{R}^{N \times L_h \times d_\text{model}}$, while a set of learnable embeddings $\mathbf{Q}_{p} \in \mathbb{R}^{N \times L_p \times d_\text{model}}$ serves as placeholders for future prediction. 
Subsequently, we concatenate these two motion-related components along the temporal dimension to form the unified trajectory sequence $\mathbf{S}_\text{traj} = [\mathbf{H}_{h}, \mathbf{Q}_{p}] \in \mathbb{R}^{N \times (L_h + L_p) \times d_\text{model}}$.
To preserve temporal order, a learnable time embedding $\mathbf{E}_\text{time}\in \mathbb{R}^{N \times (L_h+L_p) \times d_\text{model}}$ is added element-wise to $\mathbf{S}_\text{traj}$.
Finally, this time-aware trajectory sequence is concatenated with the encoded context tokens $\mathbf{E}_\text{text}, \mathbf{E}_\text{img}, \text{and } \mathbf{E}_\text{lidar}$ generated by the upstream encoders to form the unified input sequence $\mathbf{E}_\text{in}\in \mathbb{R}^{N \times (3+L_h+L_{p}) \times d_\text{model}}$ as follows:
\begin{equation}
    \mathbf{E}_\text{in} = \text{Concat}(\mathbf{S}_\text{traj} + \mathbf{E}_\text{time}, \mathbf{E}_\text{img}, \mathbf{E}_\text{lidar}, \mathbf{E}_\text{text}).
\end{equation}
This sequence is then fed into the GPT-2 backbone for autoregressive reasoning to produce the output sequence $\mathbf{E}_{out} \in \mathbb{R}^{N \times (3+L_h+L_p) \times d_\text{model}}$. Specifically, we extract the tokens corresponding to the future positions $\widehat{\mathbf{Q}}_{p} \in \mathbb{R}^{N \times L_p \times d_\text{model}}$ to serve as the learned representations for the subsequent beam prediction head.

\subsection{Beam Prediction Head}
\begin{figure}[t]
\centering
\includegraphics[width=0.98\linewidth]{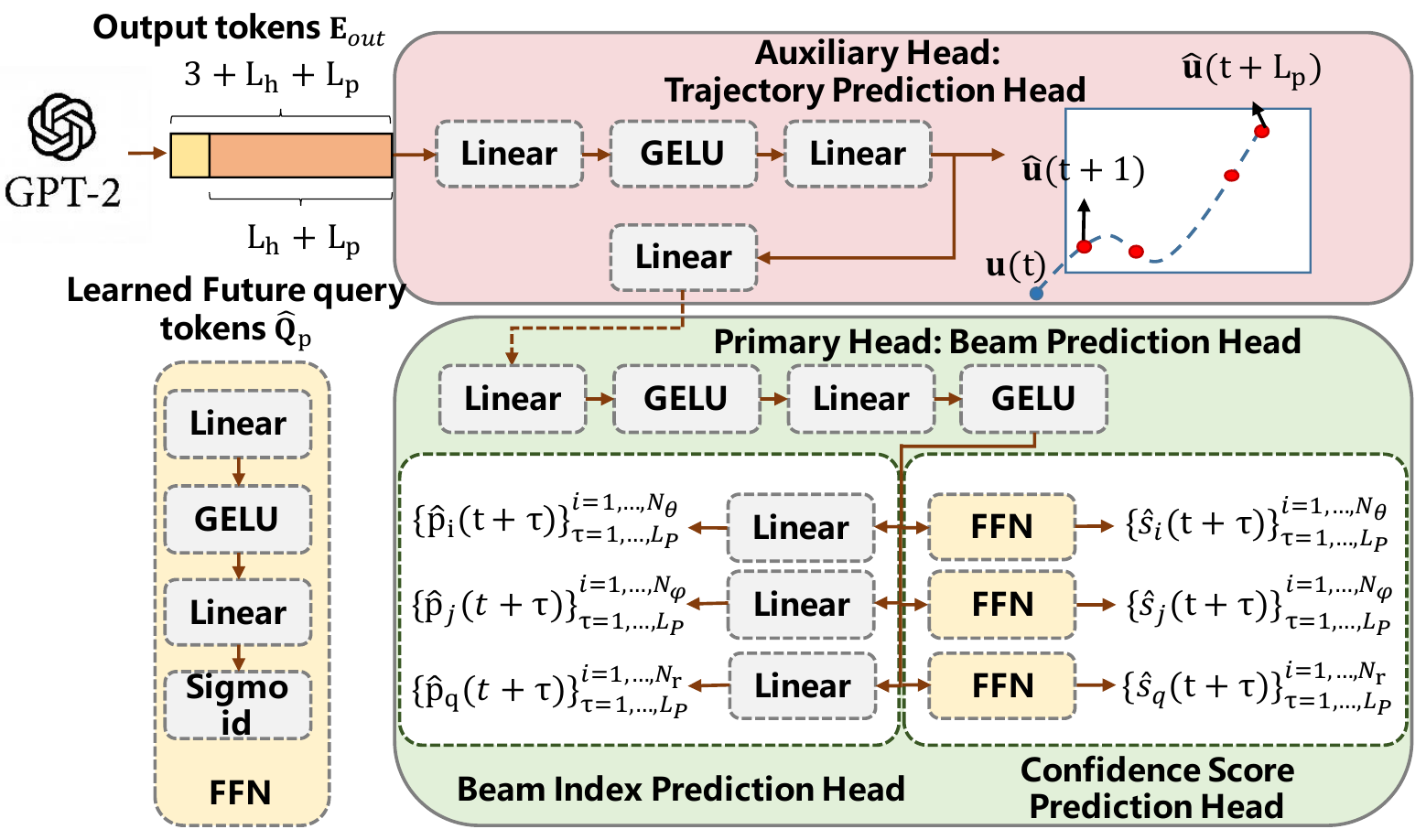}
\caption{Architecture of the designed beam prediction head.}
\label{fig:prediction_heads}
\end{figure}

As shown in Fig.~\ref{fig:prediction_heads}, the output features from the GPT-2 backbone are fed into two sequential prediction heads: an auxiliary trajectory prediction head and a decoupled near-field beam prediction head. By first utilizing the UAV's position to focus attention on the relevant surrounding environment, the model effectively narrows the candidate search space for the optimal beam.

\paragraph{Auxiliary Trajectory Prediction Head}
To facilitate spatial reasoning, we construct an auxiliary network for trajectory prediction. This module processes the learned query tokens $\widehat{\mathbf{Q}}_p$ via an MLP to regress the future 3D coordinates. The output is the predicted future trajectory sequence, denoted as $\{\widehat{\mathbf{u}}(t+\tau)\}_{\tau=1,\cdots,L_p} \in \mathbb{R}^{N \times L_p \times 3}$. 
This predicted trajectory serves as an intermediate result to assist the primary beam prediction. By explicitly recovering the UAV's future kinematic intent at each time step $t+\tau$, the network provides strong geometric priors, guiding the subsequent beam predictor to focus strictly on physically plausible locations.

\paragraph{Primary Near-Field Beam Prediction Head}

Predicting the optimal beam directly from a massive near-field codebook not only suffers from the curse of dimensionality, but also struggles with the periodic abrupt jumps inherent in 1D index labels. Because the 3D spatial parameters $(\theta, \phi, r)$ are flattened into a single 1D index sequence, physically adjacent beams frequently correspond to discontinuous index values. This misalignment destroys the intrinsic spatial correlation and motivates our design of a decoupled prediction strategy, which predicts the beam indices across each dimension independently to preserve spatial continuity.

Specifically, the output generated by the trajectory prediction head is passed through a linear projection layer to serve as the input for the beam prediction head. It then branches into three parallel streams to generate the decoupled beam probability distributions and confidence scores for the azimuth, elevation, and distance, respectively. The designed beam prediction head includes:
\begin{itemize}
    \item \textbf{Beam Index Prediction Head:} A linear classifier maps the refined features to probability distributions over the three decoupled codebook dimensions. To predict the optimal beams over the future trajectory, the network outputs probability sequences denoted as $\{\widehat{\mathbf{p}}_i(t+\tau)\}_{\tau=1,\cdots,L_p}^{i=1, \cdots, N_{\theta}} \in \mathbb{R}^{N \times L_p \times N_\theta}$, $\{\widehat{\mathbf{p}}_j(t+\tau)\}_{\tau=1,\cdots,L_p}^{j=1, \cdots, N_{\varphi}} \in \mathbb{R}^{N \times L_p \times N_\varphi}$, and $\{\widehat{\mathbf{p}}_q(t+\tau)\}_{\tau=1,\cdots,L_p}^{q=1, \cdots, N_r} \in \mathbb{R}^{N \times L_p \times N_r}$. For each future time step $t+\tau$, the final predicted sub-indices for azimuth $\hat{i}(t+\tau)$, elevation $\hat{j}(t+\tau)$, and distance $\hat{q}(t+\tau)$ are obtained by selecting the index with the maximum probability from each respective distribution.
    \item \textbf{Confidence Score Prediction Head:} Simultaneously, a feed-forward network (FFN) predicts confidence scores for the corresponding predictions. The output sequences are $\{\widehat{\mathbf{s}}_i(t+\tau)\}_{\tau=1,\cdots,L_p}^{i=1, \cdots, N_{\theta}} \in \mathbb{R}^{N \times L_p \times N_\theta}$, $\{\widehat{\mathbf{s}}_j(t+\tau)\}_{\tau=1,\cdots,L_p}^{j=1, \cdots, N_{\varphi}} \in \mathbb{R}^{N \times L_p \times N_\varphi}$, and $\{\widehat{\mathbf{s}}_q(t+\tau)\}_{\tau=1,\cdots,L_p}^{q=1, \cdots, N_r} \in \mathbb{R}^{N \times L_p \times N_r}$, where values are normalized to $[0, 1]$ via a Sigmoid function to evaluate the reliability at each time step $t+\tau$.
\end{itemize}
This decoupled design not only reduces the output space complexity from $\mathcal{O}(N_r N_\varphi N_\theta)$ to $\mathcal{O}(N_r + N_\varphi + N_\theta)$, significantly alleviating the burden of model fitting, but also allows for fine-grained control over the beam prediction accuracy.

\subsection{Adaptive Refinement}
During the inference phase, we implement an adaptive refinement post-processing strategy to mitigate unreliable predictions. Initially, we evaluate the confidence scores of the Top-1 predictions across the three decoupled dimensions (i.e., azimuth, elevation, and distance). If the confidence scores for all three dimensions simultaneously exceed a pre-defined reliability threshold $s_\text{thre}$, the Top-1 index combination is deemed reliable and directly output as the final predicted beam. However, if the confidence score of any dimension falls below $s_\text{thre}$, we trigger a localized search within a high-confidence subspace.

Specifically, we extract the Top-5 indices from the probability distribution of each dimension. The refined joint search space is denoted as $\Omega_p$, which consists of $5^3 = 125$ candidate combinations. This subspace is sufficiently small for efficient evaluation but diverse enough to encompass the optimal beam. To identify the final refined beam indices, the system evaluates all combinations in $\Omega_p$ by maximizing their joint probability:
\begin{equation}
\begin{split}
    &(\hat{i}(t+\tau), \hat{j}(t+\tau) , \hat{q}(t+\tau)) \\
    &\quad = \argmax_{(i, j, q) \in \Omega_p} \big( \widehat{\mathbf{p}}_i(t+\tau) \cdot \widehat{\mathbf{p}}_j(t+\tau) \cdot \widehat{\mathbf{p}}_q(t+\tau) \big).
\end{split}
\end{equation}
This strategy ensures that when the predicted Top-1 beam is uncertain, the model can also provide a highly reliable pool of candidates for efficient beam sweeping, thereby guaranteeing robust prediction performance.

\subsection{Training Scheme and Loss Function Design}

\paragraph{Training Scheme}
To guarantee the robustness of the proposed framework against potential sensor failures and to systematically investigate the network's performance across various modality combinations, we implement a flexible multimodal training scheme. Specifically, to simulate real-world scenarios where certain sensory inputs might be unavailable, we train the model under the following configurations:
\begin{itemize}
    \item \textbf{Support for Missing Modalities:} Our framework is designed to inherently support scenarios with incomplete sensory data. If one or more input modalities are missing or corrupted, their corresponding tokens are dynamically excluded from the input sequence $\mathbf{X}_\text{in}$, allowing the model to perform beam prediction using only the available modalities.
    \item \textbf{Parameter-Efficient Fine-Tuning:} Instead of updating all parameters of the pre-trained GPT-2, we adopt a partial fine-tuning strategy to prevent catastrophic forgetting and reduce computational cost. We freeze the majority of the transformer blocks and only update: (i) the specific projection layers of encoders and heads; (ii) the positional embeddings and LayerNorm parameters; and (iii) the top two Transformer blocks.
    \item \textbf{Output Selection:} We exclusively select the last $L_p$ output embeddings, which correspond to the learnable future query tokens $\mathbf{Q}_{p}$. These tokens are designed to aggregate global context for future inference, while the outputs associated with the preceding context and history tokens are discarded.
\end{itemize}

\paragraph{Loss Function}
The network is optimized end-to-end via the following loss function:
\begin{equation}
    \mathcal{L}_{total} = \lambda_1\mathcal{L}_\text{traj} + \lambda_2\mathcal{L}_\text{beam} + \lambda_3\mathcal{L}_\text{conf},
\end{equation}
where hyperparameters $\lambda_1,\lambda_2,\lambda_3$ balance three losses. The detailed design of the three loss terms is as follows.

First, to ensure precise intermediate localization prediction, we adopt the normalized mean square error (NMSE) averaged over all future time steps as
\begin{equation}
    \mathcal{L}_\text{traj} = \frac{1}{L_p} \sum_{\tau=1}^{L_p} \frac{\| \widehat{\mathbf{u}}(t+\tau) - \mathbf{u}(t+\tau)  \|^2}{\| \mathbf{u}(t+\tau) \|^2}.
\end{equation}

Second, we employ a soft target loss strategy to tolerate small spatial misalignments and account for the strong spatial correlation inherent in near-field beams. Instead of utilizing a rigid one-hot label, we construct smoothed target distributions $\mathbf{p}_i(t+\tau)$, $\mathbf{p}_j(t+\tau)$, and $\mathbf{p}_q(t+\tau)$ for azimuth, elevation, and distance, respectively, at each future time step $t+\tau$. Specifically, we assign fixed probability values, allocating $0.6$ to the GT beam index and $0.1$ to each of the four adjacent near-optimal indices, while setting the probabilities of all remaining indices in the codebook to zero. 

The network is then optimized to minimize the Kullback-Leibler (KL) divergence between the soft target distributions and the predicted probability distributions:
\begin{equation}
\begin{split}
    \mathcal{L}_\text{beam} &= \frac{1}{3 L_p} \sum_{\tau=1}^{L_p} \Bigg( \sum_{i=1}^{N_\theta} \mathbf{p}_i(t+\tau) \log \frac{\mathbf{p}_i(t+\tau)}{\widehat{\mathbf{p}}_i(t+\tau)} \\
    &+ \sum_{j=1}^{N_\varphi} \mathbf{p}_j(t+\tau) \log \frac{\mathbf{p}_j(t+\tau)}{\widehat{\mathbf{p}}_j(t+\tau)} \\
    &+ \sum_{q=1}^{N_r} \mathbf{p}_q(t+\tau) \log \frac{\mathbf{p}_q(t+\tau)}{\widehat{\mathbf{p}}_q(t+\tau)} \Bigg).
\end{split}
\end{equation}

Finally, the confidence score prediction is supervised by the mean squared error (MSE). To ensure that the confidence score for each dimension solely reflects its own prediction accuracy, we employ an isolation strategy for target generation. To generate the GT confidence score for the azimuth dimension at time $t+\tau$, we isolate the predicted azimuth index $\hat{i}(t+\tau)$ by pairing it with the GT elevation and distance indices, yielding the codeword as follows:
\begin{equation}
\hat{\mathbf{w}}(t+\tau) = \mathbf{w}\big(\theta_{\hat{i}(t+\tau)}, \varphi_{j^\star(t+\tau)}, r_{q^\star(t+\tau)}\big). 
\end{equation}
Utilizing the beamforming gain function defined in~\eqref{eq:beam_gain}, the target score is computed as:
\begin{equation}
    s_i(t+\tau) = \text{clamp}\left( \frac{G\big(\hat{\mathbf{w}}(t+\tau), t+\tau\big)}{G\big(\mathbf{w}^\star(t+\tau), t+\tau\big)}, 0, 1 \right),
\end{equation}
where $\mathbf{w}^\star(t+\tau)$ represents the GT codeword at the corresponding time step.
Target scores for elevation $s_j(t+\tau)$ and distance $s_q(t+\tau)$ are computed analogously by isolating their respective GT values. The final confidence loss is formulated by expanding the MSE across the three dimensions:
\begin{equation}
\begin{split}
    \mathcal{L}_\text{conf} &= \frac{1}{3 L_p} \sum_{\tau=1}^{L_p} \Bigg( \big(\widehat{\mathbf{s}}_i(t+\tau) - s_i(t+\tau)\big)^2 \\
    &\quad + \big(\widehat{\mathbf{s}}_j(t+\tau) - s_j(t+\tau)\big)^2 \\
    &\quad + \big(\widehat{\mathbf{s}}_q(t+\tau) - s_q(t+\tau)\big)^2 \Bigg).
\end{split}
\end{equation}

\section{Experimental Results} \label{sec:Experimental Results}
In this section, we first outline implementation details. We then benchmark the proposed framework against SOTA baselines, followed by in-depth ablation studies to validate the effectiveness of core components.

\subsection{Implementation Details}

\subsubsection{Dataset and Framework}
In our experiments, we utilize \textbf{Multimodal-LAE-XLMIMO}\footnote{The dataset is publicly available at: \url{https://github.com/Lmyxxn/Multimodal-NF}}, a comprehensive open-source dataset designed for multimodal sensing-aided XL-MIMO wireless communications in low-altitude scenarios. The dataset encompasses 30 diverse 3D urban environments and contains 10,770 continuous flight trajectories. For each trajectory, temporally aligned multi-modal sensory data and wireless channel features are collected over 20 consecutive time slots at the sampling interval $\Delta t=0.1$s. This rigorous collection process yields a total of 215,400 labeled samples, comprising 201,075 LoS and 14,325 NLoS samples. To evaluate the model's generalization capabilities, we adopt a scenario-based dataset splitting strategy: 22 scenes are allocated for training, with 4 reserved for validation and 4 for testing. Furthermore, to investigate the model's environmental understanding capability, we partition the test set into distinct LoS and NLoS subsets. LoS scenarios are evaluated as foundational tasks with weaker environmental dependency due to direct path visibility. Conversely, NLoS scenarios represent highly challenging cases, where the obstruction of direct signal paths necessitates high-fidelity environmental perception and complex spatial reasoning.

The XL-MIMO system operates at $f_c=7$ GHz with an $M_y \times M_z = 64 \times 64$ UPA equipped at the BS. The multi-modal inputs consist of RGB images ($224 \times 224$), LiDAR point clouds ($1024$ points/frame), and GPS coordinates corrupted by Gaussian noise with standard deviation $\sigma_{\text{GPS}} = 0.5$. The input dimension of PGA module is defined as $d_\text{in}=256$, and the unified latent dimension $d_\text{model}$ is set to 768. For temporal modeling, the model utilizes historical observations from the past $L_h=10$ time steps to predict the trajectory and optimal beams for the subsequent $L_p=10$ time steps. During the training phase, the proposed framework is fine-tuned for 100 epochs with a batch size of $N=32$, and the loss balancing hyperparameters are empirically set to $\lambda_1=0.2$, $\lambda_2=0.6$, and $\lambda_3=0.2$. The codebook resolutions are set to $N_r=10$, $N_\varphi=20$, and $N_\theta=20$. The confidence score threshold of the proposed adaptive refinement is set as $s_{\text{thre}}=0.9$.

\subsubsection{Evaluation Metrics}
To comprehensively assess the performance, we evaluate the prediction accuracy and spectral efficiency averaged over the prediction horizon $L_p$.

\begin{itemize}
    \item \textbf{Top-$K$ Accuracy:} This metric measures the probability that GT beam index is included in the set of Top-$K$ predicted candidates. We evaluate this at two granularities:
    
    (i) \textit{Decomposed Accuracy}: This evaluates the prediction performance of azimuth, elevation, and distance independently. The Top-K accuracies for the three dimensions are defined as:
    \begin{equation}
    \begin{split}
        \text{Acc}_{\text{Top}K}^{i} &= \frac{1}{L_p} \sum_{\tau=1}^{L_p} \mathbb{I} \Big( i^\star(t+\tau) \in \mathcal{I}_{\text{Top-}K}(t+\tau) \Big), \\
        \text{Acc}_{\text{Top-}K}^{j} &= \frac{1}{L_p} \sum_{\tau=1}^{L_p} \mathbb{I} \Big( j^\star(t+\tau) \in \mathcal{J}_{\text{Top-}K}(t+\tau) \Big), \\
        \text{Acc}_{\text{Top-}K}^{q} &= \frac{1}{L_p} \sum_{\tau=1}^{L_p} \mathbb{I} \Big( q^\star(t+\tau) \in \mathcal{Q}_{\text{Top-}K}(t+\tau) \Big),
    \end{split}
    \end{equation}
    where $\mathcal{I}_{\text{top-}K}$, $\mathcal{J}_{\text{top-}K}$, and $\mathcal{Q}_{\text{top-}K}$ denote the sets of $K$ indices with the highest probabilities for azimuth, elevation, and distance, respectively.

    (ii) \textit{joint Accuracy}: This evaluates the success of the overall beam index tuple prediction. It is defined as:
    \begin{equation}
        \text{Acc}_{\text{top-}K}^{\text{joint}} = \frac{1}{L_p} \sum_{\tau=1}^{L_p} \mathbb{I} \Big( k^\star(t+\tau) \in \Omega_{\text{top-}K}(t+\tau) \Big),
    \end{equation}
    where $\Omega_{\text{top-}K}$ is the candidate set containing the $K$ overall indices with the highest joint probabilities.

    \item \textbf{Average Achievable Rate:} Following the achievable rate defined in~\eqref{eq:achievable_rate}, the average rate over the prediction horizon is expressed as:
    \begin{equation}
        R_a = \frac{1}{L_p} \sum_{\tau=1}^{L_p} R\big(\hat{\mathbf{w}}(t+\tau), t+\tau\big),
    \end{equation}
where $\hat{\mathbf{w}}(t+\tau) \in \mathcal{W}$ denotes the beam codeword constructed from the predicted indices $\big(\hat{i}(t+\tau), \hat{j}(t+\tau), \hat{q}(t+\tau)\big)$ at future time step $t+\tau$. 
    \item \textbf{Average Normalized Beamforming Gain:} This metric evaluates the gap between the predicted beam and the optimal beam. Using the beamforming gain defined in~\eqref{eq:beam_gain}, the average normalized beamforming gain over the prediction horizon is calculated as:
    \begin{equation}
        \bar{G} = \frac{1}{L_p} \sum_{\tau=1}^{L_p} \frac{G\big(\hat{\mathbf{w}}(t+\tau), t+\tau\big)}{G\big(\mathbf{w}^\star(t+\tau), t+\tau\big)}.
    \end{equation}

     \item \textbf{Trajectory Prediction Mean Absolute Error (MAE)}
We evaluate the precision of the intermediate output predicted trajectory utilizing
\begin{equation}
    \text{MAE} = \frac{1}{L_p} \sum_{\tau=1}^{L_p} \| \mathbf{u}(t+\tau) - \widehat{\mathbf{u}}(t+\tau) \|.
\end{equation}

\end{itemize}

\subsubsection{Baselines and Ablation Studies}
To comprehensively evaluate the proposed framework, we benchmark its performance against two categories of representative algorithms and conduct ablation studies to validate our design:
\begin{itemize}
    \item \textbf{Deep Learning (DL)-based Sequence Models:} We first benchmark against lightweight and widely adopted sequence models as baselines, specifically RNN~\cite{khunteta2021recurrent} and LSTM~\cite{shah2022multi} with history GPS positions as the input. Furthermore, we compare our framework against M2BeamLLM~\cite{zheng2025m2beamllm}, a SOTA multi-modal LLM-driven method.
    \item \textbf{Efficient Near-Field Beam Training Algorithms:} We also compare against near-field search methods, specifically Hierarchical Search~\cite{lu2024hierarchical} and Two-stage Search~\cite{wu2024two}. For a fair comparison, the pilot overhead for these search-based baselines is strictly limited to match the average pilot budget consumed by our adaptive refinement phase.
    \item \textbf{Ablation Studies:} We conduct comprehensive ablation studies to evaluate the core contributions of our proposed framework, structured into two main parts. Part I investigates the impact of varying combinations of input modalities and demonstrates the significant performance gains achieved by the proposed adaptive refinement strategy. Part II assesses the effectiveness of the proposed core components by isolating the LLM backbone, the decoupled beam index prediction head, the auxiliary trajectory prediction head, and the designed textual prompt.
\end{itemize}

\newcommand{\imgwidth}{0.98\linewidth}
\newcommand{\imgvspace}{-0.4em}       

\begin{figure}[t]
\centering
\subfigure[Top-1 accuracy]{%
    \includegraphics[width=\imgwidth]{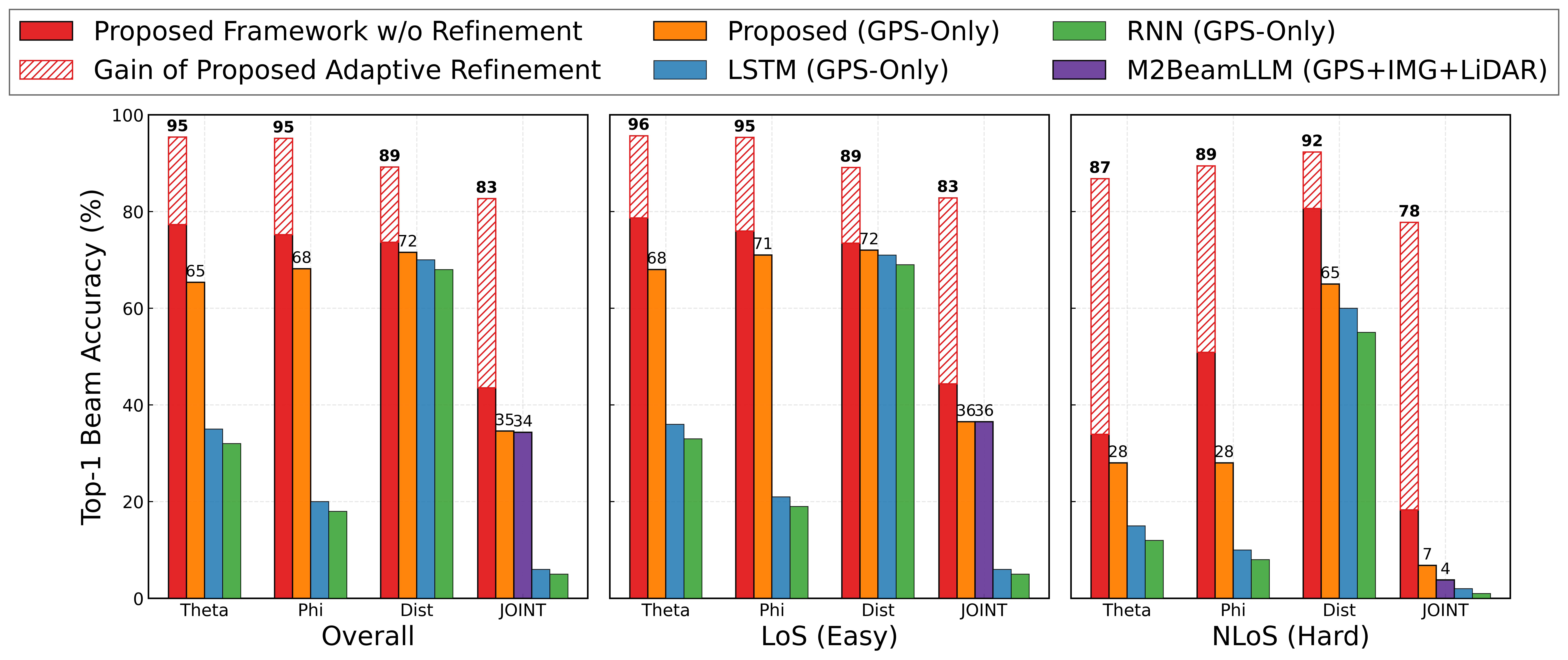}%
}\vspace{\imgvspace}

\subfigure[Top-5 accuracy]{%
    \includegraphics[width=\imgwidth]{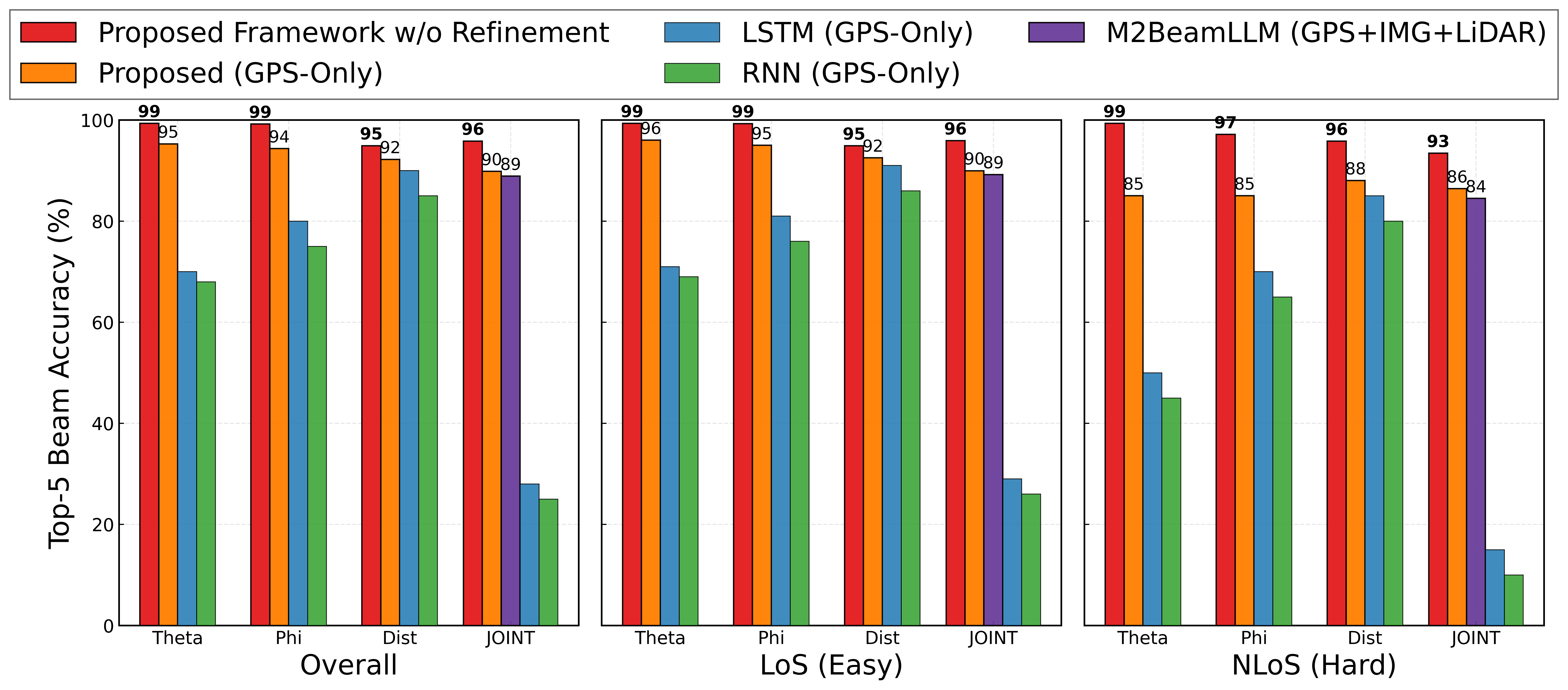}%
}

\caption{Beam prediction accuracy comparison for different deep learning-based algorithms across overall, LoS, and NLoS scenarios. (a) Top-1 accuracy. (b) Top-5 accuracy.}
\label{fig:beamacc}
\end{figure}

\subsection{Beam Prediction Performance}

\subsubsection{Accuracy Comparison with DL Baselines}
As illustrated in Fig.~\ref{fig:beamacc}(a), the proposed framework without adaptive refinement surpasses all other baselines in all the accuracy metrics, including $ \text{Acc}_{\text{Top}1}^{i}, \text{Acc}_{\text{Top}1}^{j}, \text{Acc}_{\text{Top}1}^{q}, \text{Acc}_{\text{Top}1}^{\text{joint}}$ and $\text{Acc}_{\text{Top}5}^{i}, \text{Acc}_{\text{Top}5}^{j}, \text{Acc}_{\text{Top}5}^{q}, \text{Acc}_{\text{Top}5}^{\text{joint}}$. Moreover, the proposed framework with GPS-only inputs achieves a Top-1 joint accuracy of 35\% across all test scenarios. This performance significantly exceeds that of traditional sequence models such as RNN~\cite{khunteta2021recurrent} and LSTM~\cite{shah2022multi}, which struggle to surpass the 10\% threshold given the same input. Notably, it even slightly outperforms M2BeamLLM~\cite{zheng2025m2beamllm}, which utilizes GPS, images, and LiDAR data as the inputs. This superiority mainly stems from the proposed structure-aware beam prediction strategy, which effectively simplifies the high-dimensional near-field search space compared to the overall beam index classification approach of M2BeamLLM and uses auxiliary trajectory prediction to further enhance the accuracy.
Building upon this superior architecture, the proposed confidence-score-based adaptive refinement mechanism further improves the reliability and accuracy of the prediction. Specifically, it boosts the Top-1 joint beam prediction accuracy to 83\% across all test scenarios. Even in the most difficult NLoS environments, the refinement mechanism successfully elevates the accuracy from 18\% to 78\%, ensuring highly reliable beam alignment where conventional baselines completely collapse.

\begin{figure}[t]
\centering
\includegraphics[width=0.9\linewidth]{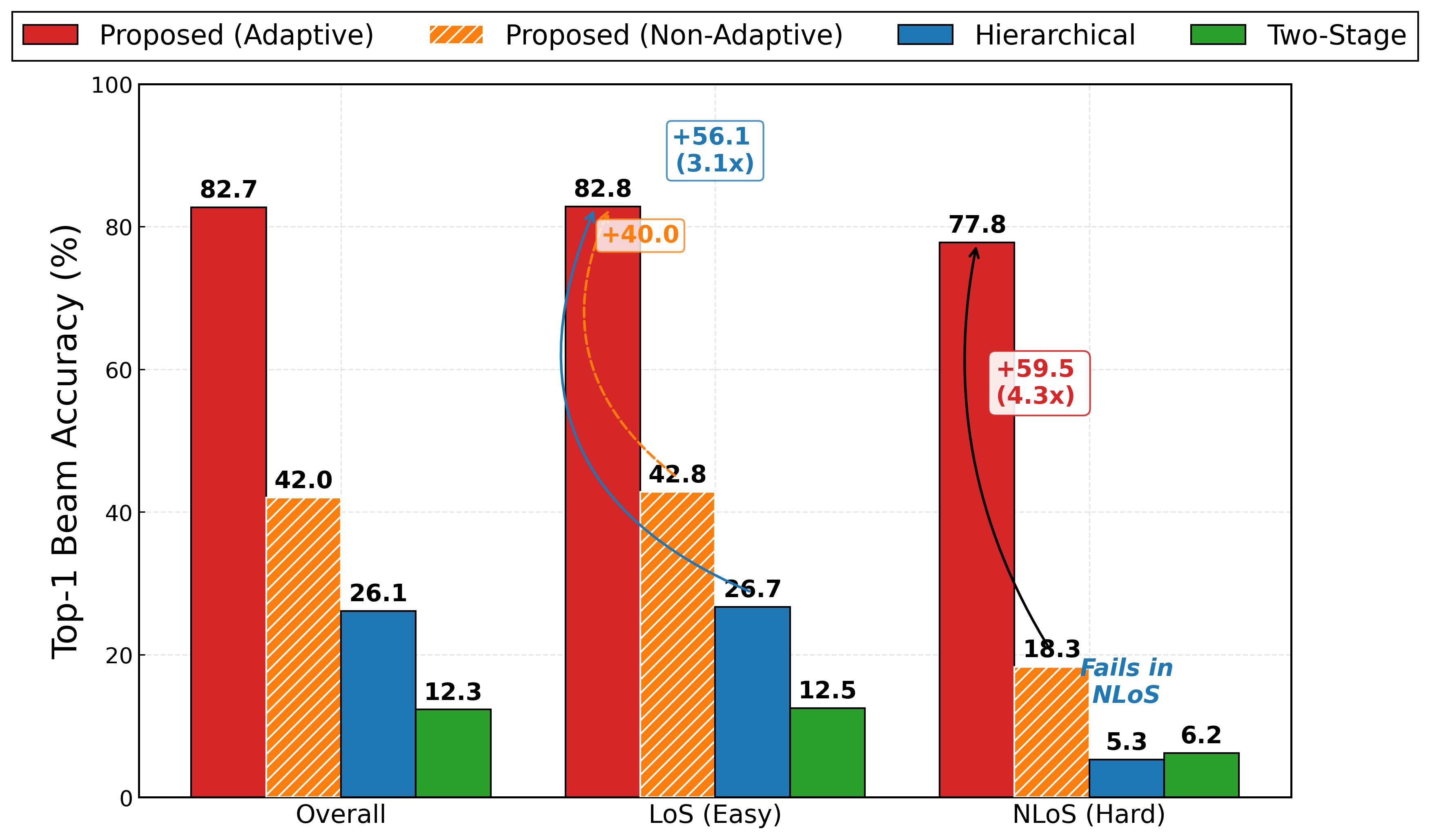}
\caption{Performance comparison of Top-1 beam prediction accuracy against near-field baselines with consistent beam training overhead.}
\label{fig:beamacc_nf}
\end{figure}

\begin{figure*}[t]
\centering
\includegraphics[width=0.88\linewidth]{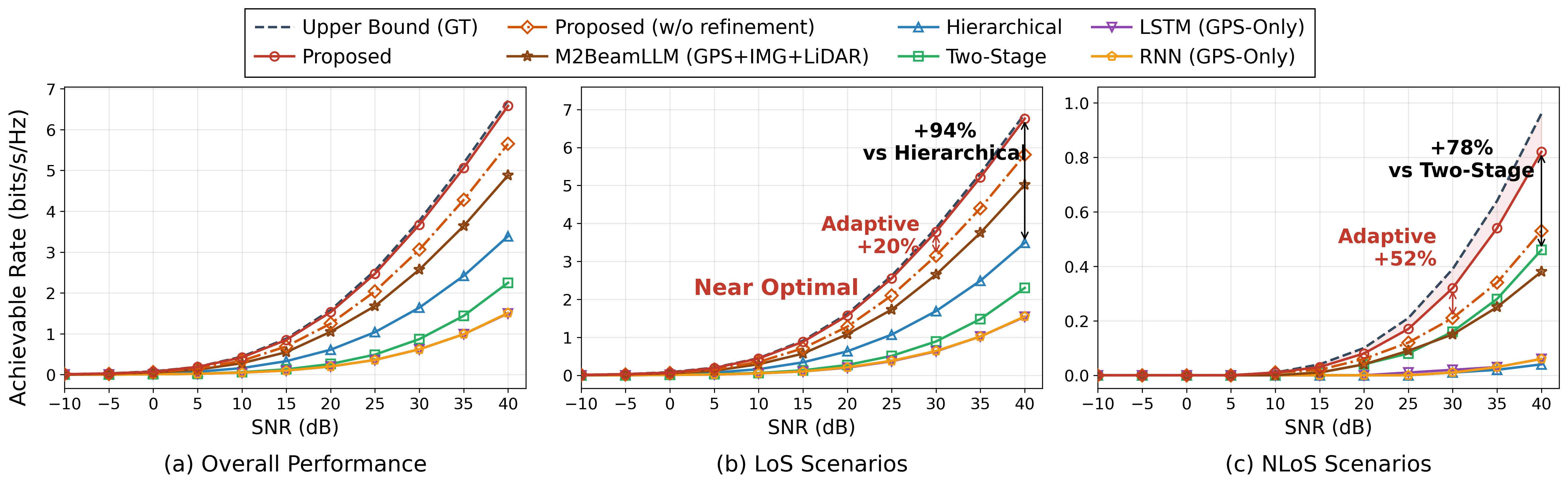}
\caption{Achievable rate comparison against baselines and the GT upper bound with consistent beam training overhead.}
\label{fig:rate_snr}
\end{figure*}

As shown in Fig.~\ref{fig:beamacc}(b), the Top-5 joint accuracy of the proposed framework exceeds 90\% in both LoS and NLoS scenarios, while the accuracy for each decomposed index (azimuth, elevation, and distance) consistently surpasses 95\%, which exceeds other baselines. Such high Top-5 performance validates the reliability of the generated candidate pool, providing a solid foundation for the subsequent adaptive refinement stage to achieve accurate alignment with low overhead.

\subsubsection{Accuracy Comparison with Near-Field Beam Training Baselines}
Since conventional beam training methods are designed to output the optimal beam index, we first evaluate the Top-1 joint accuracy $\text{Acc}_{\text{Top}1}^{\text{joint}}$ for comparison.  Fig.~\ref{fig:beamacc_nf} illustrates the performance attained under a fixed overhead of 90, matching the average overhead incurred by the proposed adaptive refinement strategy, which operates at a 90\% confidence threshold and 71.6\% needs to sweep from the candidate pool (pool size = 125). Under this overhead budget, conventional baselines suffer from severe under-sampling, yielding at most 26.1\% overall accuracy. In contrast, our framework achieves a robust 82.7\% accuracy. Specifically, in LoS scenarios, our method outperforms Hierarchical Search~\cite{lu2024hierarchical} by 3.1 times. In challenging NLoS scenarios where beam training baselines fundamentally fail, our adaptive refinement proves critical, boosting accuracy by 4.3 times over the framework without adaptive refinement.

\begin{table*}[t]
\centering
\caption{Performance Comparison and Ablation Study of the Proposed LLM-Driven Multi-Modal Framework.}
\label{tab:ablation_study}
\renewcommand{\arraystretch}{1.15}
\setlength{\tabcolsep}{4pt}

\begin{tabular}{llccccc}
\toprule
\textbf{Configuration} & \textbf{Scenario} & \textbf{Pos MAE [m]} $\downarrow$ & \textbf{Top-1 Acc [\%]} $\uparrow$ & \textbf{Top-5 Acc [\%]} $\uparrow$ & \textbf{Norm. Gain} $\uparrow$ & \textbf{High Conf. [\%]}$^{\dagger}$ $\uparrow$ \\
\midrule

\multicolumn{7}{c}{\textbf{Part I: Ablation of Input Modalities}} \\
\midrule

\multirow{3}{*}{\shortstack[l]{\textbf{Full Modalities} \\ \textbf{w/ Adaptive Refinement}}} 
 & \textbf{Overall} 
    & \textbf{0.8959} 
    & \textbf{82.66} \textcolor{blue}{\scriptsize \textbf{(+39.58)}} 
    & \textbf{95.82} 
    & \textbf{0.9462} \textcolor{blue}{\scriptsize \textbf{(+0.2172)}} 
    & \textbf{29.4} \\
 & \textbf{LoS}     
    & \textbf{0.9032} 
    & \textbf{82.82} \textcolor{blue}{\scriptsize \textbf{(+38.94)}} 
    & \textbf{95.90} 
    & \textbf{0.9490} \textcolor{blue}{\scriptsize \textbf{(+0.2052)}} 
    & \textbf{30.0} \\
 & \textbf{NLoS}    
    & \textbf{0.6675} 
    & \textbf{77.75} \textcolor{blue}{\scriptsize \textbf{(+59.91)}} 
    & \textbf{93.43} 
    & \textbf{0.8558} \textcolor{blue}{\scriptsize \textbf{(+0.5923)}} 
    & \textbf{13.2} \\
\cmidrule(lr){2-7}

\multirow{3}{*}{\shortstack[l]{Full Modalities\\ \textit{(w/o Adaptive Refinement)}}} 
 & Overall & 0.8959 & 43.08 & 95.82 & 0.7290 & 29.4 \\
 & LoS      & 0.9032 & 43.88 & 95.90 & 0.7438 & 30.0 \\
 & NLoS    & 0.6675 & 17.84 & 93.43 & 0.2635 & 13.2 \\
\cmidrule(lr){2-7}

\multirow{3}{*}{GPS + IMG + Prompt} 
 & Overall & 1.0337 & 42.47 & 93.40 & 0.7130 & 24.7 \\
 & LoS      & 1.0414 & 43.36 & 93.49 & 0.7276 & 25.5 \\
 & NLoS    & 0.7933 & 14.32 & 90.75 & 0.2526 & 1.5  \\
\cmidrule(lr){2-7}
\multirow{3}{*}{GPS + LiDAR + Prompt} 
 & Overall & 0.9955 & 42.47 & 92.83 & 0.7123 & 22.0 \\
 & LoS      & 1.0019 & 43.52 & 92.94 & 0.7286 & 22.6 \\
 & NLoS    & 0.7939 & 9.47  & 89.43 & 0.1990 & 4.0  \\
\cmidrule(lr){2-7}
\multirow{3}{*}{GPS + Prompt} 
 & Overall & 1.2346 & 37.93 & 91.64 & 0.6485 & 16.9 \\
 & LoS      & 1.2473 & 38.89 & 91.81 & 0.6636 & 17.4 \\
 & NLoS    & 0.8333 & 7.93  & 89.29 & 0.1737 & 0.0  \\
\cmidrule(lr){2-7}
\multirow{3}{*}{GPS Only} 
 & Overall & 1.3117 & 35.58 & 89.84 & 0.6201 & 15.3 \\
 & LoS      & 1.3250 & 36.50 & 89.95 & 0.6350 & 15.8 \\
 & NLoS    & 0.8950 & 6.80  & 86.40 & 0.1510 & 0.0  \\
\midrule
\midrule

\multicolumn{7}{c}{\textbf{Part II: Ablation of the Proposed Framework Components}} \\
\midrule

\multirow{3}{*}{\shortstack[l]{Full Modalities with LSTM\\ \textit{(w/o LLM Backbone)}}} 
 & Overall & 9.8036 & 6.70 & 29.57 & 0.1396 & -- \\
 & LoS      & 9.9179 & 6.88 & 30.32 & 0.1430 & -- \\
 & NLoS    & 6.2022 & 1.10 & 6.17 & 0.0302 & -- \\
\cmidrule(lr){2-7}

\multirow{3}{*}{\shortstack[l]{Full Modalities \\ \textit{(w/o Decoupled Head)}}} 
 & Overall & 1.1500 & 36.80 & 90.50 & 0.6200 & 22.0 \\
 & LoS     & 1.1580 & 37.60 & 90.70 & 0.6350 & 22.5 \\
 & NLoS    & 0.9000 & 11.80 & 84.85 & 0.1510 & 6.4  \\
\cmidrule(lr){2-7}

\multirow{3}{*}{\shortstack[l]{Full Modalities \\ \textit{(w/o Auxiliary Head)}}} 
 & Overall & -- & 38.70 & 91.99 & 0.6681 & 26.0 \\
 & LoS      & -- & 39.45 & 92.10 & 0.6825 & 26.5 \\
 & NLoS    & -- & 15.20 & 88.50 & 0.2150 & 11.4 \\
\cmidrule(lr){2-7}

\multirow{3}{*}{\shortstack[l]{Full Modalities \\ \textit{(w/o Textual Prompt)}}} 
 & Overall & 1.1041 & 41.80 & 92.92 & 0.7144 & 19.5 \\
 & LoS      & 1.1096 & 42.91 & 93.00 & 0.7327 & 20.0 \\
 & NLoS    & 0.9317 & 6.61  & 90.31 & 0.1373 & 4.4  \\

\bottomrule
\multicolumn{7}{l}{\footnotesize $^{\dagger}$ High Conf. represents the percentage of predictions with a confidence score exceeding 0.9.}
\end{tabular}
\end{table*}

\begin{figure*}[t]
\centering
\includegraphics[width=0.8\linewidth]{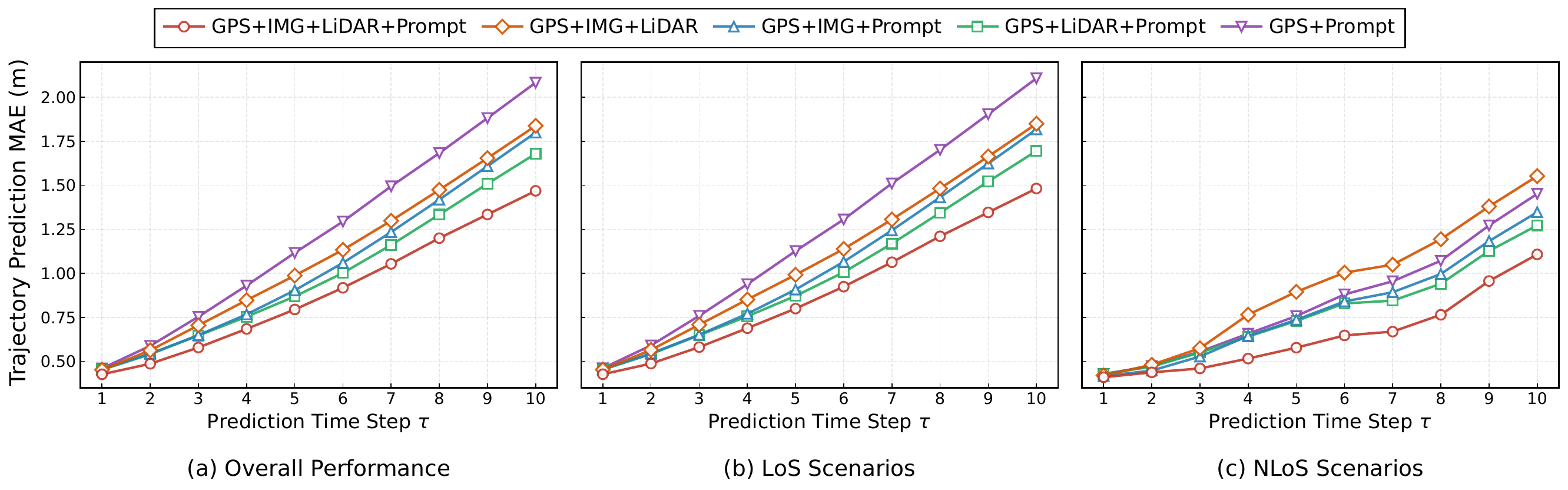}
\caption{Ablation study on the proposed auxiliary trajectory prediction head: Trajectory MAE versus prediction time step $\tau$ under various input modality combinations.}
\label{fig:tra_acc}
\end{figure*}

\begin{figure*}[t]
\centering
\includegraphics[width=0.65\linewidth]{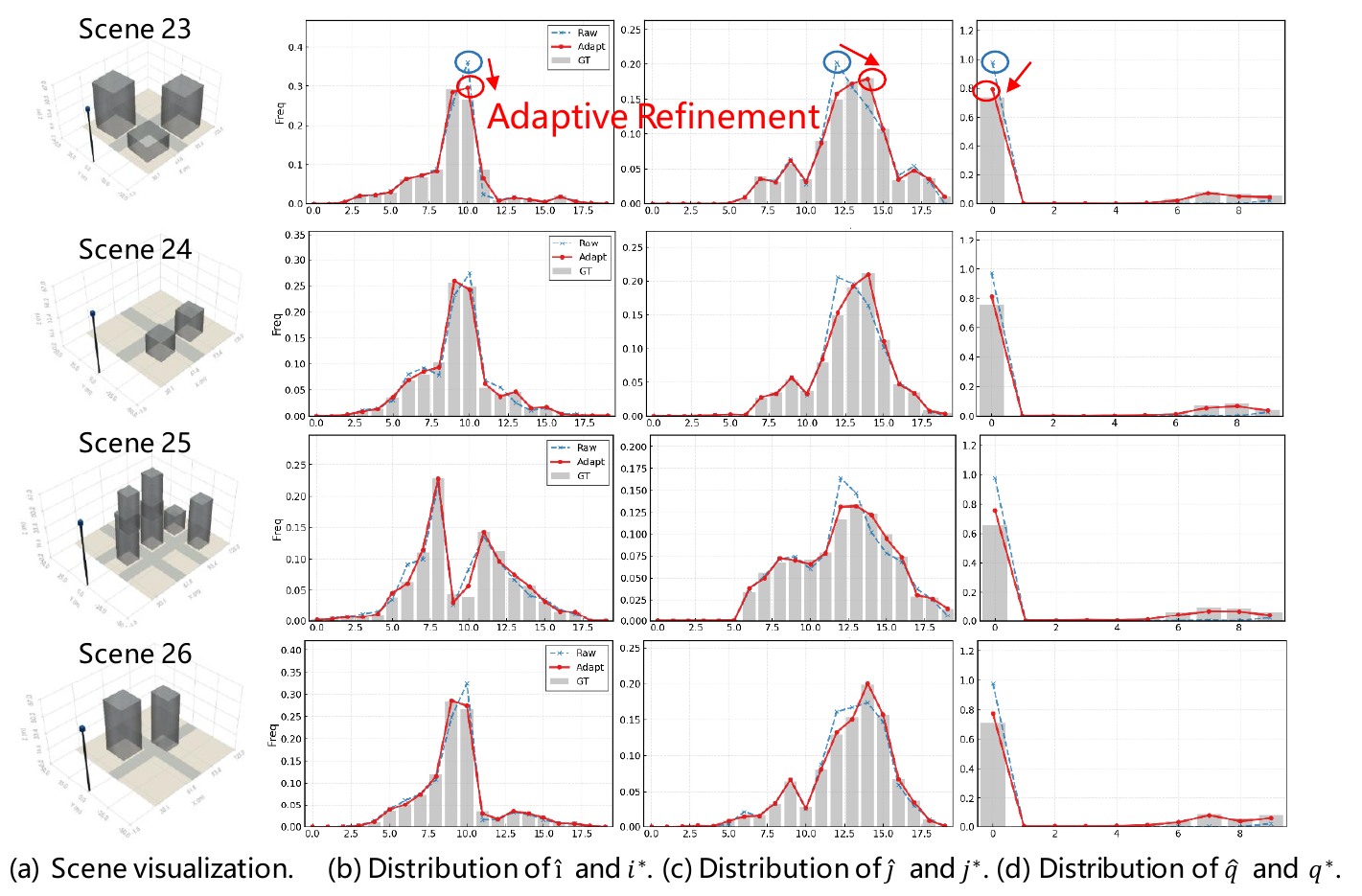}
\caption{Comparison of the predicted beam index distributions of the test scenes for the proposed framework with and without adaptive refinement, alongside the GT distribution.}
\label{fig:beam_index_dist}
\end{figure*}

\subsubsection{System Achievable Rate Comparison}

Fig.~\ref{fig:rate_snr} compares the system achievable rate of the proposed framework against various baselines and the GT upper bound. Compared to DL-based prediction models (RNN~\cite{khunteta2021recurrent}, LSTM~\cite{shah2022multi}, and M2BeamLLM~\cite{zheng2025m2beamllm}), our framework demonstrates superior performance gain across all test scenarios, which is particularly pronounced in NLoS environments (Fig.~\ref{fig:rate_snr}(c)), where the substantial gap between our method and other models underscores the superior environmental perception and spatial reasoning capabilities of the proposed framework. 
Furthermore, when compared to efficient near-field beam training baselines (Hierarchical~\cite{lu2024hierarchical} and Two-stage~\cite{wu2024two}) under the same overhead, our approach maintains a near-optimal rate that closely tracks the GT upper bound. In LoS scenarios, our framework outperforms the Hierarchical baseline by a staggering 94\%. Finally, the confidence-score-based adaptive refinement proves essential for robustness. It brings a 20\% and 52\% rate gain over the proposed framework without refinement in LoS and NLoS scenarios, respectively. Notably, in NLoS cases, the adaptive mechanism effectively bridges the performance gap, achieving a 78\% higher rate than the Two-stage baseline and ensuring reliable connectivity in complex environments.

\subsection{Ablation Study}
To evaluate the effectiveness of the various components of the proposed framework, we conduct a comprehensive ablation study as summarized in Table~\ref{tab:ablation_study}. The analysis is divided into three aspects, including the impact of input modalities, the improvements brought by the adaptive refinement strategy, and the contribution of individual framework components.

\subsubsection{Impact of Input Modalities}
As shown in Part I of Table~\ref{tab:ablation_study}, the \textbf{full modalities} configuration (GPS+IMG+LiDAR+Prompt) achieves the best performance across all metrics, yielding the lowest positioning MAE of 0.8959 m and an initial Top-1 accuracy of 43.08\% (without adaptive refinement). 
By comparing different modality subsets, it can be observed that the integration of visual semantics and depth information is crucial for spatial awareness. For example, transitioning from ``GPS-Only'' to the full modalities setup reduces the positioning MAE by 31.7\% (1.3117 m to 0.8959 m), introduces a 7.5\% Top-1 accuracy gain, and boosts the high confidence ratio from 15.3\% to 29.4\%. 
Fig.~\ref{fig:tra_acc} also presents an ablation study on the proposed auxiliary trajectory prediction head, illustrating the trajectory MAE versus the prediction time step $\tau$ under various input modality combinations. The results demonstrate that the proposed framework achieves high-precision trajectory tracking concurrently with robust beam prediction. Notably, the full modalities input exhibits superior stability, maintaining the positioning error within a small range of 0.3~m to 1.5~m. In contrast, incomplete configurations suffer from performance degradation as the prediction horizon extends. This comparison validates that, alongside the dominant positional data, the integration of diverse modalities is indispensable for accurately capturing complex motion dynamics and environmental contexts. 
Notably, Table~\ref{tab:ablation_study} shows a lower trajectory prediction MAE in NLoS scenarios compared to LoS. This is largely because NLoS conditions are frequently in low-speed trajectory mode or hovering near blockages. 

\subsubsection{Improvements of the Proposed Adaptive Refinement}
As shown in Part I of Table~\ref{tab:ablation_study}, the proposed \textbf{adaptive refinement strategy} provides a decisive performance leap. In  overall test scenarios, it surges the Top-1 accuracy from 43.08\% to 82.66\%. This improvement is even more significant in NLoS cases, where the accuracy climbs from 17.84\% to 77.75\%, demonstrating the framework's ability to correct prediction errors through the proposed confidence-score-based adaptive refinement.
 Fig.~\ref{fig:beam_index_dist} further compares the predicted beam index distributions of our framework (with and without adaptive refinement) against the GT distribution across all test scenarios. The histograms denote the GT, while the curves represent the raw and refined predictions. The high degree of overlap between the final predictions and the GT demonstrates the robustness of the proposed framework. The result also validates the efficacy of the proposed confidence-score-based adaptive refinement. While raw predictions without refinement occasionally show minor deviations in peak positioning or magnitude caused by inherent model uncertainties, the refinement mechanism effectively corrects these discrepancies, resulting in highly accurate beam predictions.

\subsubsection{Effectiveness of Proposed Framework Components}
Part II of Table~\ref{tab:ablation_study} validates the effectiveness of our core architectural designs, including:

\begin{itemize}
    \item \textbf{LLM Backbone:} Replacing the utilized GPT-2 backbone with a conventional sequence model, i.e., an LSTM with the same structure as~\cite{shah2022multi}, results in a total performance collapse. Specifically, the trajectory prediction MAE increases to 9.8036 m and the Top-1 accuracy drops to 6.70\%. This underscores the LLM's superior capability in processing heterogeneous multi-modal sequences and reasoning and comprehending complex environmental mappings.
    \item \textbf{Decoupled Beam Index Prediction Head:} Ablating the proposed decoupled beam prediction head and predicting the overall beam index directly leads to a severe performance degradation. Specifically, the Overall Top-1 accuracy drops to 36.80\%, and the NLoS normalized beamforming gain decreases to 0.1510. This demonstrates that decomposing the massive near-field codebook space into distance ($r$) and angular ($\varphi, \theta$) dimensions effectively mitigates the curse of dimensionality inherent in large-scale classifications and improves the accuracy.
    \item \textbf{Auxiliary Trajectory Prediction Head:} Ablating the proposed auxiliary trajectory prediction head results in a noticeable performance drop, with the Top-1 accuracy decreasing from 43.08\% to 38.70\%. This demonstrates that trajectory prediction acts as an effective prior to guide beam prediction, thereby elevating the model's understanding of the environment.
    \item \textbf{Textual Prompt:} Ablating the proposed textual prompt degrades overall performance. Notably, it causes a severe drop in the NLoS Top-1 accuracy (from 17.84\% to 6.61\%) and a significant reduction in the LoS high confidence ratio (from 30.0\% to 20.0\%). It indicates that the designed textual prompts not only secure the prediction confidence in LoS scenarios but also provide the indispensable reasoning capability required to overcome complex NLoS blockages.
\end{itemize}

\section{Conclusion}
\label{sec:Conclusion}
This paper proposed a structure-aware multimodal LLM framework to tackle the inherent inefficiency of near-field XL-MIMO beam training in complex 3D environments. By integrating GPS data, RGB images, and LiDAR data, the proposed framework leverages the emergent reasoning capabilities of LLMs to achieve a profound understanding of the coupling between near-field beams and physical surroundings. To circumvent the curse of dimensionality in the joint angular-distance domain, we implemented a structure-aware beam prediction strategy that mirrors the 3D geometric structure of the codebook, further enhanced by an auxiliary trajectory prediction head for spatial guidance. Moreover, a trustworthy adaptive refinement mechanism was introduced to dynamically trigger small-scale scanning based on confidence scores, achieving the trade-off between alignment accuracy and pilot overhead. Extensive experimental results demonstrate that our framework significantly outperforms state-of-the-art baselines in both LoS and NLoS scenarios, underscoring the potential of multimodal LLMs for reliable near-field communications in 6G and beyond.

\appendices
\section{Position-Guided Aggregation Mechanism}
\label{appendix:pga}
The PGA module operates via a cross-attention mechanism. For a given sensory modality, it treats the UAV position $\mathbf{u}(t) \in \mathbb{R}^{N \times 1 \times 3}$ as the query, and the raw feature map $\mathbf{F} \in \mathbb{R}^{N \times N_\text{token} \times d_\text{in}}$ as both the key and value. The aggregation is formulated as:
\begin{equation}
    \mathbf{E} = \text{Softmax}\left( \frac{(\mathbf{u}(t)\mathbf{W}_Q)(\mathbf{F}\mathbf{W}_K)^T}{\sqrt{d_\text{model}}} + \mathbf{M} \right) (\mathbf{F}\mathbf{W}_V),
\end{equation}
where $\mathbf{W}_Q \in \mathbb{R}^{3 \times d_\text{model}}$ and $\mathbf{W}_K, \mathbf{W}_V \in \mathbb{R}^{d_\text{in} \times d_\text{model}}$ are learnable projections. The term $\mathbf{M} \in \mathbb{R}^{N \times 1 \times N_\text{token}}$ is the spatial bias matrix, and $d_\text{model}$ is the unified latent dimension. The resulting context token is $\mathbf{E} \in \mathbb{R}^{N \times 1 \times d_\text{model}}$.

 \small\bibliographystyle{./bibliography/IEEEtran}
 \bibliography{ref}

@article{wang2024xlmimo,
  author    = {Z. Wang and others},
  title     = {A tutorial on extremely large-scale {MIMO} for {6G}: Fundamentals, signal processing, and applications},
  journal   = {IEEE Commun. Surv. Tut.},
  volume    = {26},
  number    = {3},
  year      = {2024},
  month     = {Jul.},
  pages     = {1560--1605}
}

@article{han2023toward,
  author    = {Y. Han and others},
  title     = {Toward extra large-scale {MIMO}: New channel properties and low-cost designs},
  journal   = {IEEE Internet Things J.},
  volume    = {10},
  number    = {16},
  year      = {2023},
  month     = {Aug.},
  pages     = {14569--14594}
}

@article{qi2020hierarchical,
  author    = {C. Qi and others},
  title     = {Hierarchical codebook-based multiuser beam training for millimeter wave massive {MIMO}},
  journal   = {IEEE Trans. Wireless Commun.},
  volume    = {19},
  number    = {12},
  year      = {2020},
  month     = {Dec.},
  pages     = {8142--8152}
}

@article{liu2025near,
  author    = {Y. Liu and others},
  title     = {Near-field communications: A comprehensive survey},
  journal   = {IEEE Commun. Surv. Tut.},
  volume    = {27},
  number    = {3},
  year      = {2025},
  month     = {Jun.},
  pages     = {1687--1728}
}

@article{cui2022channel,
  author    = {M. Cui and L. Dai},
  title     = {Channel estimation for extremely large-scale {MIMO}: Far-field or near-field?},
  journal   = {IEEE Trans. Commun.},
  volume    = {70},
  number    = {4},
  year      = {2022},
  month     = {Apr.},
  pages     = {2663--2677}
}

@article{luo2024efficient,
  author    = {J. Luo and others},
  title     = {Efficient hybrid near- and far-field beam training for {XL-MIMO} communications},
  journal   = {IEEE Trans. Veh. Technol.},
  volume    = {73},
  number    = {12},
  year      = {2024},
  month     = {Dec.},
  pages     = {19785--19790},
  doi       = {10.1109/TVT.2024.3447129}
}

@article{xu2025near,
  author    = {Z. Xu and others},
  title     = {Near-optimal near-field beam training: From searching to inference},
  journal   = {IEEE Trans. Wireless Commun.},
  volume    = {24},
  number    = {11},
  year      = {2025},
  month     = {Nov.},
  pages     = {9173--9185},
  doi       = {10.1109/TWC.2025.3571488}
}

@article{xue2024ai,
  author    = {Q. Xue and others},
  title     = {{AI/ML} for beam management in {5G-Advanced}: A standardization perspective},
  journal   = {IEEE Veh. Technol. Mag.},
  volume    = {19},
  number    = {4},
  year      = {2024},
  month     = {Dec.},
  pages     = {64--72}
}

@article{lin2025bridge,
  author    = {X. Lin},
  title     = {The bridge toward {6G}: {5G-Advanced} evolution in {3GPP} Release 19},
  journal   = {IEEE Commun. Standard Mag.},
  volume    = {9},
  number    = {1},
  year      = {2025},
  month     = {Mar.},
  pages     = {28--35}
}

@article{cui2023near,
  author    = {M. Cui and L. Dai},
  title     = {Near-field wideband channel estimation for extremely large-scale {MIMO}},
  journal   = {Sci. China Inf. Sci.},
  volume    = {66},
  number    = {7},
  year      = {2023},
  month     = {Jun.},
  pages     = {172303}
}

@article{li2025keypoint,
  author    = {M. Li and others},
  title     = {Keypoint detection empowered near-field user localization and channel reconstruction},
  journal   = {IEEE Trans. Wireless Commun.},
  volume    = {24},
  number    = {7},
  year      = {2025},
  month     = {Jul.},
  pages     = {5664--5677},
  doi       = {10.1109/TWC.2025.3548626}
}

@article{liu2025large,
  author    = {W. Liu and others},
  title     = {Large-model {AI} for near field beam prediction: A {CNN-GPT2} framework for {6G} {XL-MIMO}},
  journal   = {arXiv preprint arXiv:2510.22557},
  year      = {2025},
  month     = {Oct.}
}

@inproceedings{charan2022vision,
  author    = {G. Charan and others},
  title     = {Vision-position multi-modal beam prediction using real millimeter wave datasets},
  booktitle = {Proc. IEEE Wireless Commun. Netw. Conf. (WCNC)},
  year      = {2022},
  month     = {Apr.},
  address   = {Austin, TX, USA},
  pages     = {2727--2731},
  doi       = {10.1109/WCNC51071.2022.9771835}
}

@article{chen2023mmwave,
  author    = {L. Chen and others},
  title     = {MmWave beam tracking with spatial information based on extended Kalman filter},
  journal   = {IEEE Wireless Commun. Lett.},
  volume    = {12},
  number    = {4},
  year      = {2023},
  month     = {Apr.},
  pages     = {615--619}
}

@article{jayaprakasam2017robust,
  author    = {S. Jayaprakasam and others},
  title     = {Robust beam-tracking for mmWave mobile communications},
  journal   = {IEEE Commun. Lett.},
  volume    = {21},
  number    = {12},
  year      = {2017},
  month     = {Dec.},
  pages     = {2654--2657}
}

@article{lu2024hierarchical,
  author    = {Y. Lu and others},
  title     = {Hierarchical beam training for extremely large-scale {MIMO}: From far-field to near-field},
  journal   = {IEEE Trans. Commun.},
  volume    = {72},
  number    = {4},
  year      = {2024},
  month     = {Apr.},
  pages     = {2247--2259}
}

@article{wu2024two,
  author    = {C. Wu and others},
  title     = {Two-stage hierarchical beam training for near-field communications},
  journal   = {IEEE Trans. Veh. Technol.},
  volume    = {73},
  number    = {2},
  year      = {2024},
  month     = {Feb.},
  pages     = {2032--2044}
}

@inproceedings{wu2024near,
  author    = {X. Wu and others},
  title     = {Near-field beam training with {DFT} codebook},
  booktitle = {Proc. IEEE Wireless Commun. Netw. Conf. (WCNC)},
  year      = {2024},
  month     = {Apr.},
  address   = {Dubai, United Arab Emirates},
  pages     = {1--6},
  doi       = {10.1109/WCNC57260.2024.10571054}
}

@inproceedings{shokri2015beam,
  author    = {H. S. Ghadikolaei and others},
  title     = {Beam-searching and transmission scheduling in millimeter wave communications},
  booktitle = {Proc. IEEE Int. Conf. Commun. (ICC)},
  year      = {2015},
  month     = {Jun.},
  pages     = {1292--1297}
}

@inproceedings{yaman2016reducing,
  author    = {Y. Yaman and P. Spasojevic},
  title     = {Reducing the {LOS} ray beamforming setup time for {IEEE} 802.11ad and {IEEE} 802.15.3c},
  booktitle = {Proc. IEEE Mil. Commun. Conf. (MILCOM)},
  year      = {2016},
  month     = {Nov.},
  pages     = {448--453}
}

@inproceedings{shen2018mobility,
  author    = {L.-H. Shen and others},
  title     = {Mobility-aware fast beam training scheme for {IEEE} 802.11ad/ay wireless systems},
  booktitle = {Proc. IEEE Wireless Commun. Netw. Conf. (WCNC)},
  year      = {2018},
  month     = {Apr.},
  pages     = {1--6}
}

@inproceedings{khunteta2021recurrent,
  author    = {S. Khunteta and A. K. R. Chavva},
  title     = {Recurrent neural network based beam prediction for millimeter-wave {5G} systems},
  booktitle = {Proc. IEEE Wireless Commun. Netw. Conf. (WCNC)},
  year      = {2021},
  month     = {Mar.},
  pages     = {1--6},
  doi       = {10.1109/WCNC49053.2021.9417573}
}

@article{shah2022multi,
  author    = {S. H. A. Shah and S. Rangan},
  title     = {Multi-cell multi-beam prediction using auto-encoder {LSTM} for mmWave systems},
  journal   = {IEEE Trans. Wireless Commun.},
  volume    = {21},
  number    = {12},
  year      = {2022},
  month     = {Dec.},
  pages     = {10366--10380},
  doi       = {10.1109/TWC.2022.3183603}
}

@article{jiang2023lidar,
  author    = {S. Jiang and others},
  title     = {{LiDAR} aided future beam prediction in real-world millimeter wave {V2I} communications},
  journal   = {IEEE Wireless Commun. Lett.},
  volume    = {12},
  number    = {2},
  year      = {2023},
  month     = {Feb.},
  pages     = {212--216},
  doi       = {10.1109/LWC.2022.3220556}
}

@article{charan2024camera,
  author    = {G. Charan and others},
  title     = {Camera based mmWave beam prediction: Towards multi-candidate real-world scenarios},
  journal   = {IEEE Trans. Veh. Technol.},
  volume    = {74},
  number    = {4},
  year      = {2024},
  month     = {Apr.},
  pages     = {5897--5913}
}

@article{zhao2025multi,
  author    = {Y. Zhao and others},
  title     = {Multi-modal large models based beam prediction: An example empowered by {DeepSeek}},
  journal   = {arXiv preprint arXiv:2506.05921},
  year      = {2025}
}

@article{zheng2025m2beamllm,
  author    = {C. Zheng and others},
  title     = {{M2BeamLLM}: Multimodal sensing-empowered mmWave beam prediction with large language models},
  journal   = {arXiv preprint arXiv:2506.14532},
  year      = {2025}
}

@article{sheng2025beam,
  author    = {Y. Sheng and others},
  title     = {Beam prediction based on large language models},
  journal   = {IEEE Wireless Commun. Lett.},
  volume    = {14},
  number    = {5},
  year      = {2025},
  month     = {May},
  pages     = {1406--1410},
  doi       = {10.1109/LWC.2025.3526188}
}

@article{radford2019language,
  author    = {A. Radford and others},
  title     = {Language models are unsupervised multitask learners},
  journal   = {OpenAI blog},
  volume    = {1},
  number    = {8},
  pages     = {9},
  year      = {2019}
}

@article{chen2025janus,
  author    = {X. Chen and others},
  title     = {{Janus-Pro}: Unified multimodal understanding and generation with data and model scaling},
  journal   = {arXiv preprint arXiv:2501.17811},
  year      = {2025}
}

@inproceedings{hu2022lora,
  author    = {E. J. Hu and others},
  title     = {{LoRA}: Low-rank adaptation of large language models},
  booktitle = {Proc. Int. Conf. Learn. Represent. (ICLR)},
  year      = {2022}
}

@article{alkhateeb2023deepsense,
  author    = {A. Alkhateeb and others},
  title     = {{DeepSense} {6G}: A large-scale real-world multi-modal sensing and communication dataset},
  journal   = {IEEE Commun. Mag.},
  volume    = {61},
  number    = {9},
  year      = {2023},
  month     = {Sept.},
  pages     = {122--128},
  doi       = {10.1109/MCOM.001.2200734}
}

@article{mao2025multimodal,
  author    = {T. Mao and others},
  title     = {{Multimodal-Wireless}: A large-scale dataset for sensing and communication},
  journal   = {arXiv preprint arXiv:2511.03220},
  year      = {2025}
}

@inproceedings{sionnaRT,
  author    = {J. Hoydis and others},
  title     = {{Sionna RT}: Differentiable ray tracing for radio propagation modeling},
  booktitle = {Proc. IEEE Globecom Workshops (GC Wkshps)},
  address   = {Kuala Lumpur, Malaysia},
  year      = {2023},
  month     = {Dec.},
  pages     = {317--321}
}

@inproceedings{devlin2019bert,
  author    = {J. Devlin and others},
  title     = {{BERT}: Pre-training of deep bidirectional transformers for language understanding},
  booktitle = {Proc. NAACL-HLT},
  address   = {Minneapolis, MN, USA},
  year      = {2019},
  month     = {Jun.},
  pages     = {4171--4186}
}

@inproceedings{he2016deep,
  author    = {K. He and others},
  title     = {Deep residual learning for image recognition},
  booktitle = {Proc. IEEE CVPR},
  address   = {Las Vegas, NV, USA},
  year      = {2016},
  month     = {Jun.},
  pages     = {770--778}
}

@inproceedings{qi2017pointnet,
  author    = {C. Qi and others},
  title     = {{PointNet}: Deep learning on point sets for {3D} classification and segmentation},
  booktitle = {Proc. IEEE CVPR},
  address   = {Honolulu, HI, USA},
  year      = {2017},
  month     = {Jul.},
  pages     = {652--660}
}

\vspace{12pt}
\color{red}

\end{document}